\def\paperversion{final}
\pdfoutput=1
\def\anonymoussubmission{1} %
\newif\ifIsArxivEnv\IsArxivEnvtrue
\newif\ifshellescapeavailable\shellescapeavailablefalse
\newif\ifIsAcmart\IsAcmarttrue
\def\texmacrosdir{tex-macros/}

\newif\ifsinglecolumn\singlecolumnfalse
\newif\ifwidemargins\widemarginsfalse
\newif\ifwarning\warningfalse
\newif\ifshowcomments\showcommentsfalse
\newif\ifblinded\blindedfalse
\newif\ifreport\reportfalse
\newif\ifcopyrightspace\copyrightspacefalse
\newif\ifacknowledgments\acknowledgmentsfalse
\newif\ifshowpagenumbers\showpagenumberstrue
\newif\iffinalformat\finalformatfalse
\newif\ifweb\webfalse

\ifx\anonymoussubmission\undefined
\else
  \if\anonymoussubmission 1
    \blindedtrue
  \else
    \blindedfalse
  \fi
\fi
\IfFileExists{.blinded}{\blindedtrue}

\ifx\paperversion\xxxxundefined
\PackageError{paperversions}{*** No valid document version was specified.
Macro paperversions must be be defined as one of (markup, draft,
local, submission, final, web, tr, trdraft)}
\fi

\def\xxversion{\csname xx\paperversion\endcsname}
\newif\ifsawversion\sawversionfalse

\ifcase\xxversion\relax
    \widemarginstrue
    \singlecolumntrue
    \warningtrue
    \sawversiontrue
\or %draft
    \warningtrue
    \showcommentstrue
    \sawversiontrue
\or %local
    \warningtrue
    \showcommentsfalse
    \blindedfalse
    \sawversiontrue
    \acknowledgmentstrue
\or %submission
    \sawversiontrue
\or %final
    \blindedfalse
    \sawversiontrue
    \copyrightspacetrue
    \acknowledgmentstrue
    \showpagenumbersfalse
    \finalformattrue
\or %tr
    \singlecolumntrue
    \blindedfalse
    \sawversiontrue
    \reporttrue
    \acknowledgmentstrue
    \webtrue
\or %trdraft
    \blindedfalse
    \singlecolumntrue
    \showcommentstrue
    \sawversiontrue
    \reporttrue
\or %finaldraft
    \blindedfalse
    \showcommentstrue
    \copyrightspacetrue
    \acknowledgmentstrue
    \sawversiontrue
    \finalformattrue
\or %web
    \blindedfalse
    \sawversiontrue
    \copyrightspacetrue
    \acknowledgmentstrue
    \finalformattrue
    \webtrue
\or %blindtr
    \singlecolumntrue
    \blindedtrue
    \acknowledgmentsfalse
    \sawversiontrue
    \reporttrue
    \webtrue
\fi

\ifsawversion
\else
\fi

\let\xxversion=\undefined

\ifblinded
	\ifIsAcmart
		\documentclass[sigconf, review, anonymous, urlbreakonhyphens, nonacm]{acmart}
	\else
		\documentclass[]{interact}
	\fi
\else
	\ifIsAcmart
		\documentclass[sigconf, urlbreakonhyphens, nonacm]{acmart} %
	\else
		\documentclass[]{interact}
	\fi
\fi

\usepackage{xspace}

\usepackage{graphicx,eso-pic}
\usepackage{amsmath, stmaryrd}
\usepackage{\texmacrosdir utf8math}
\usepackage{caption,subcaption}

\usepackage{\texmacrosdir ttquot}

\usepackage[normalem]{ulem}

\ifIsAcmart
\else
	\usepackage{xurl}
	\usepackage[natbibapa,nodoi]{apacite}
\fi

\ifshowcomments
	\usepackage[inline]{\texmacrosdir authcomments}
\else
	\usepackage[disabled]{\texmacrosdir authcomments}
\fi

\newcommenter{OA}{0,1,0}
\newcommenter{HZ}{0.5,0,1}

\usepackage{listings}
\ifIsArxivEnv
	\ifshellescapeavailable
		\usepackage[finalizecache, outputdir=figures-minted]{minted}
	\else
		\usepackage[frozencache, outputdir=figures-minted]{minted}
	\fi
\else
	\usepackage{minted}
\fi

\def\BibTeX{{\rm B\kern-.05em{\sc i\kern-.025em b}\kern-.08em
	T\kern-.1667em\lower.7ex\hbox{E}\kern-.125emX}}

\title{Secure Distributed Applications the \textsc{Decent} Way}

\ifblinded
	\author{Anonymized for review}
\else

	\ifIsAcmart
		\author{Haofan Zheng}
			\orcid{0000-0001-6831-8574}
			\email{hzheng6@ucsc.edu}
			\affiliation{%
				\institution{UC Santa Cruz}
				\city{Santa Cruz}
				\state{CA}
				\postcode{95064}
			}

			\author{Owen Arden}
			\email{owen@soe.ucsc.edu}
			\affiliation{%
				\institution{UC Santa Cruz}
				\city{Santa Cruz}
				\state{CA}
				\postcode{95064}
			}
	\else
		\author{
			\name{Haofan Zheng (hzheng6@ucsc.edu) and Owen Arden (owen@soe.ucsc.edu)}
			\affil{%
				UC Santa Cruz, Santa Cruz, CA, USA
			}
		}
	\fi

\fi

\ifIsAcmart
\fi

\begin{document}
	\newcommand{\DECENT}{Decent\xspace}
\newcommand{\DecentSvr}{\DECENT Server\xspace}
\newcommand{\DecentComp}{\DECENT Component\xspace}
\newcommand{\DecentApp}{\DECENT App\xspace}
\newcommand{\DecentVrfy}{\DECENT Verifier\xspace}
\newcommand{\DecentRevc}{\DECENT Revoker\xspace}
\newcommand{\AuthList}{AuthList\xspace}
\newcommand{\CppCodeInline}[1]{\mintinline[breaklines,breakanywhere]{c++}{#1}}

\newcommand{\dComp}{component\xspace}
\newcommand{\dSvr}{server\xspace}
\newcommand{\dApp}{app\xspace}
\newcommand{\dVrfy}{verifier\xspace}
\newcommand{\dRevc}{revoker\xspace}

\newcommand{\DComp}{Component\xspace}
\newcommand{\DVrfy}{Verifier\xspace}
\newcommand{\DRevc}{Revoker\xspace}

\ifIsAcmart
\else
	\renewcommand{\cite}[1]{\citep{#1}}
\fi

\setcounter{secnumdepth}{4}

\ifIsAcmart
\else
	\maketitle
\fi

\ifIsAcmart



\fi

\begin{abstract}
	
Remote attestation (RA) authenticates code running in trusted
execution environments (TEEs), allowing trusted code to be deployed even on
untrusted hosts.  However, trust relationships established by one
component in a distributed application may impact the security of
other components, making it difficult to reason about
the security of the application as a whole.  Furthermore, traditional
RA approaches interact badly with modern web service design, which
tends to employ small interacting microservices, short session
lifetimes, and little or no state.

This paper presents the Decent Application Platform, a framework for
building secure decentralized applications. Decent applications
authenticate and authorize distributed enclave components using a protocol based on
\emph{self-attestation certificates}, a reusable credential based on
RA and verifiable by a third party.  Components mutually authenticate each 
other
not only based on their code, but also based on the other components
they trust, ensuring that no transitively-connected components receive
unauthorized information.
While some other TEE frameworks support mutual authentication in some form, 
Decent is the only system that supports 
mutual authentication without requiring an additional trusted third party besides the 
trusted hardware's manufacturer.
We have verified the secrecy and authenticity
of Decent application data in ProVerif, and
implemented two applications to evaluate Decent's expressiveness and performance:
DecentRide, a ride-sharing service, and DecentHT, a distributed hash table.
On the YCSB benchmark, we show that DecentHT achieves 7.5x higher throughput
and 3.67x lower latency compared to a non-Decent implementation.

\end{abstract}

\ifIsAcmart
\else
\fi

\ifIsAcmart
	\maketitle
\else
\fi

\section{Introduction} \label{sec:Intro}

A fundamental challenge in building secure decentralized applications
is that untrustworthy nodes may arbitrarily deviate from their
expected behavior.  A remote service may appear to execute a
well-known system component when in fact it is executing a maliciously
modified version.  Trusted execution environments (TEEs) partially address this
challenge by shifting trust from the host executing the component to
the manufacturer of the host's hardware.  By provisioning unique
private keys to each TEE and certifying the corresponding public keys,
third parties can authenticate messages produced within a genuine TEE
as long as the key is kept secret.\footnote{Doing so is not
trivial: implementation errors, side channels, and
physical attacks could potentially leak these keys. We assume the security of
TEE platforms for this paper, although that is demonstrably
untrue for Intel SGX~(e.g., \cite{foreshadow}).}
Of particular
interest are messages that attest to what code is executing within the
TEE. These messages, called remote attestations (RAs), allow a remote node
to prove it is executing an authentic system component.

Authenticating a TEE requires some degree of trust in a centralized
entity such as the chip manufacturer. Once the TEE platform is authenticated,
deciding whether to permit the TEE to access protected
resources is determined by the entities that control those resources.
A malicious application running within an authentic TEE should never be given
access to secret data since the TEE does not prevent it from disclosing secrets
to untrusted parties.

In a decentralized TEE application, mutually distrustful entities may wish to
protect their resources within the same application, and may disagree on which
entities are trusted.
Conceptually, RA places \emph{trust in code} at the center of
a distributed application's security. Rather than consider whether a
host will execute a component faithfully, developers can focus on the
intrinsic behavior of the component to ensure their application
behaves as expected.

\begin{figure}
	\centering

	\begin{subfigure}[t]{0.5\linewidth}
		\centering
		\includegraphics[width=0.9\linewidth]{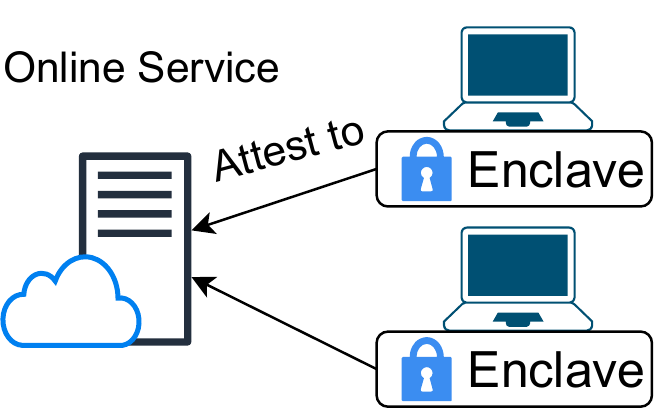}
		\caption{Enclaves attest to a server.}
		\label{sfig:Intro-CltAtt2Svr}
	\end{subfigure}%
	~
	\begin{subfigure}[t]{0.5\linewidth}
		\centering
		\includegraphics[width=0.9\linewidth]{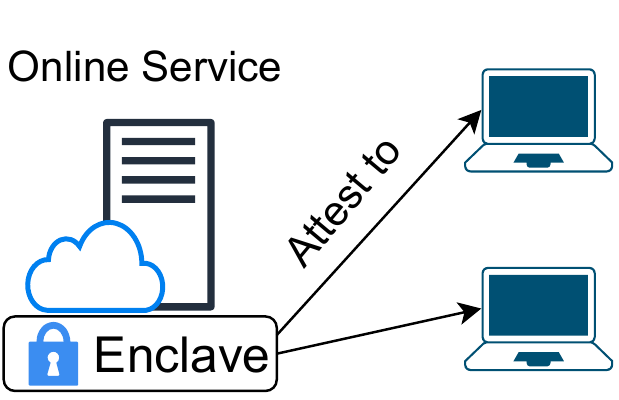}
		\caption{Enclave attests to clients.}
		\label{sfig:Intro-SvrAtt2Clt}
	\end{subfigure}

	\caption{Traditional RA authentication}
	\label{fig:single}
\end{figure}

Current TEE frameworks force programmers to work at the wrong level of
abstraction where they must deal with many low-level protocol details.
These frameworks work fine for simple scenarios where one host wishes
to authenticate remote code running on another host, such as when a
server wishes to authenticate client code, as illustrated in
Figure~\ref{sfig:Intro-CltAtt2Svr}, or a client wishes to
authenticate server code, as illustrated in Figure~\ref{sfig:Intro-SvrAtt2Clt}.
To authenticate a server component, the server attests to a
cryptographic hash of the code it is executing, signs it with its
private keys and sends it to the client. The client verifies the
signature to ensure the message originated from an authentic TEE and
compares the hash to an expected value.

\begin{figure}
	\centering
	\includegraphics[width=0.5\linewidth]{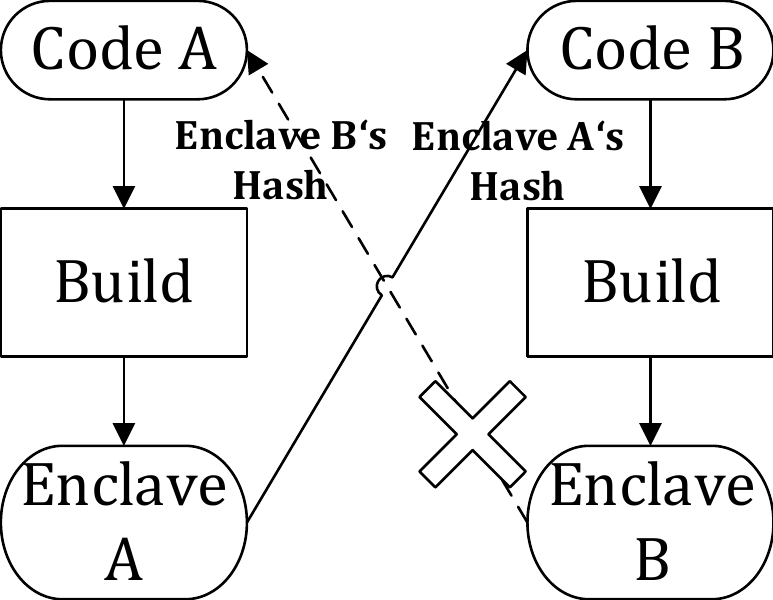}
	\caption{Mutual RA authentication. Enclave A and B cannot hardcode
                      each other's identity directly in their code since it creates a circular dependency.}
	\label{fig:mutual}
\end{figure}

Even modest extensions of this scenario introduce challenges.
Consider the scenario, illustrated in Figure~\ref{fig:mutual}, where
two components wish to mutually authenticate each other.  Each TEE
attests to a cryptographic hash of the code it is executing and sends
the signed attestation to the other host.  Authenticating one
component to the other is subtly different than
Figure~\ref{fig:single} because the verifying component must know
what hash value to expect from the remote host. Otherwise,
the component will be unable to distinguish authorized components from unauthorized ones.
Hardcoding this value in the verifying component is not possible for
both components because of a circular dependency: the hash of one
component depends on the hash of the other component.

If we exclude expected code hashes when determining the hash used to
identify a component, then each component must obtain the expected
hashes at runtime.  An honest component must therefore have a way of
authenticating the hashes to prevent attackers (such as the
component's host) from introducing malicious components in place of the honest
component's dependencies.

Addressing this circular dependency forces many systems (e.g.,
\cite{hunt2018ryoan,russinovich2019ccf,sun2019mutualauth,shinde2017panoply})
to introduce trusted third-parties to sign binaries or configurations
to prevent malicious hosts from subverting applications by introducing
a malicious component.  We are unaware of prior work that solves the
mutual authentication problem in its general form.  Beekman et
al.~\cite{beekman16at} propose a work-around that combines components
into a single binary that is running in different modes for each
component.  This method clearly does not scale to applications with many components,
and may not even be practical for moderately sized components
if the memory available to the TEE is limited.\footnote{Intel SGX currently limits
the size of the Enclave Page Cache (where enclave binaries are loaded) to about 90MB of
usable space.}

\begin{figure}
	\centering
	\includegraphics[width=0.9\linewidth]{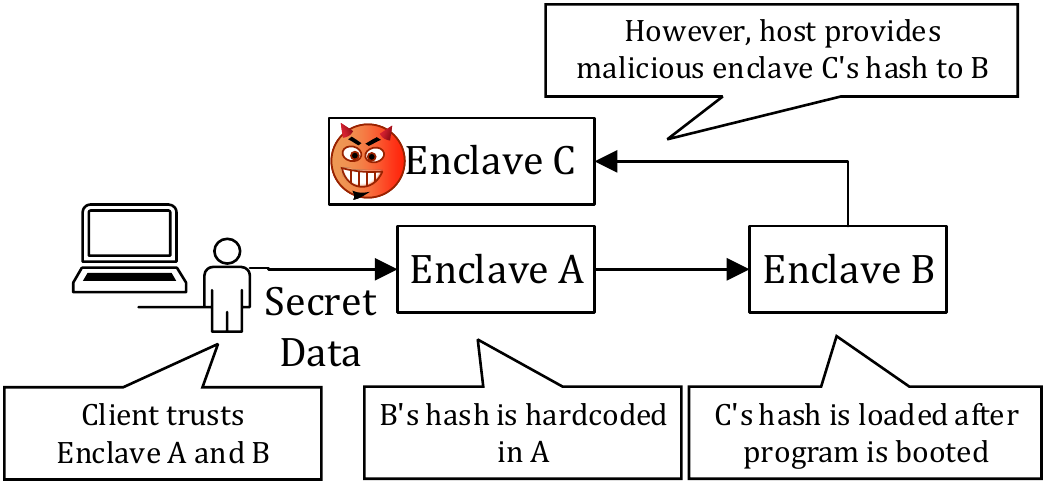}
	\caption{Allowing hosts to authorize enclaves independently could permit flows to malicious enclaves.}
	\label{fig:trans}
\end{figure}

Even if one component has hardcoded hashes, if any component it
(transitively) depends on loads hashes dynamically, its security could be
compromised.
For example, in Figure~\ref{fig:trans}, suppose
component "A" authenticates "B" against a hash of its code.
Since "B"'s hash is fixed in "A"'s code, a malicious host cannot
substitute a malicious version of "B".  However, suppose
"B" loads the expected hash of "C" dynamically.
Then if "B"'s host provides the hash of a malicious component for "C", "B" may leak
"A"'s messages to "C".  The core problem is that even though "A" authenticated "B"'s code,
it could not authenticate which components "B" would trust.

Since it is only possible to hardcode component hashes that exist at
compile time, most TEE applications with multiple components face some
version of the above problems. %
If a trusted third party exists, such as a universally-trusted developer,
the problem is easily solved: components could accept
expected hash values that are signed by the developer.
Note that because information may propagate through multiple
components as in Figure~\ref{fig:trans}, the party must be trusted by
all components in the system. Such a highly trusted entity may
not always exist.\footnote{Strictly speaking,
  all entities in a system secured by TEEs must trust the TEE manufacturer.
In this paper, we assume that entities trust the TEE and its manufacturer for
authentication, but not authorization.
That is, entities accept an attestation as proof
of what component is running in a remote TEE, but do not rely on the manufacturer to
determine which components are trustworthy.
A malicious manufacturer (or catastrophic flaw in its implementation)
could subvert the attestation process to authenticate unauthorized TEEs as
authorized ones, but we consider such attacks outside the scope of this paper.
}  Even so, were this entity to be compromised, it would
result in catastrophic failure of the security of the system since the
entity's credentials could arbitrarily change the components of the
system.
To ensure the end-to-end security of distributed and decentralized applications,
components need more flexible and expressive mechanisms for
authorization that do not require universally-trusted entities to enforce.

Another challenge for decentralized TEE applications arises when components
are replicated for scalability.  For example, serverless
applications such as those built on AWS Lambda \cite{awslambda},
Google Cloud Functions \cite{gcloudfunc}, or Azure Functions \cite{azurefunc},
reduce resource costs by factoring their
program logic into stateless components that interact simply with
persistent data storage.  If demand for the component suddenly
increases, new replicas are launched to meet the demand.  If demand
drops off, replicas may be killed to reclaim their resources.
Therefore, to take full advantage of serverless platforms, components
need to have relatively low startup costs and be able to process
requests from any client, even if a different replica previously
processed requests from the same client.

RA composes poorly with serverless design in part
because an attestation only authenticates a specific replica.
Whenever a client is presented with a new replica, the attestation
protocol must be repeated and authenticated.  Repeated attestations
can introduce significant latency.
For example, the standard Intel SGX EPID-based RA protocol
requires a client and server to exchange at least five
messages, and additionally requires communication with the
Intel Attestation Service (IAS) to verify the attestation report
\cite{intel2018ra,anati2013epid}{}.\footnote{Intel
also supports DCAP~\cite{scarlata2018dcap}, which reduces
interactions with IAS by deploying a custom report generating enclave.  This
enclave must be authenticated by IAS via a special certification enclave,
but verifiers can use the IAS root certificate to verify attestation reports without contacting IAS.
We discuss DCAP and relate it to our Decent Server in §\ref{sec:SelfAttest}.}

Distributed TEE applications frequently
amortize this cost by agreeing on some cheaper, ephemeral means of
authenticating future communication such as message authentication
codes (MACs) based on a shared key.  Unfortunately, since the cost of RA
is so high, the cost is fully amortized only for long sessions with
many requests.  Since most serverless-style applications frequently
spawn new replicas and have relatively short sessions, reducing the
overhead of authenticating new replicas could significantly improve
the performance of applications that use RA.

Finally, the software development lifecycle also presents challenges
for decentralized TEE applications. RA allows
developers (and indirectly, users) to specify how
binaries may interact, but code changes frequently over their lifetime.
Updating one component should rarely
result in downtime for the entire system.  Therefore, developers need a
mechanism to securely authorize new components and revoke old components
even after a system is deployed.

To address these challenges we have developed the \DECENT Application
Platform, a framework for building secure decentralized applications
using TEEs.
The major contribution of this paper is introducing a framework that enables mutual
authentication of dynamic sets of authorized enclave components in a
decentralized, distributed \emph{application instance} without requiring additional trusted
third parties (other than the hardware manufacturer).
Instead of forcing developers to build ad-hoc
authorization mechanisms on top of RA, \DECENT
developers refer to the \dComp{s} their application depends on using
a high-level service name.  At deployment, \DECENT nodes specify an
\emph{authorization list}, or \AuthList , that defines the \dComp{s} that are
authorized to implement each service.  At runtime, the \DECENT platform
authenticates each remote \dComp and ensures it is authorized to
perform the desired service.

\DECENT ensures that malicious hosts cannot compromise the
confidentiality or integrity of a \DECENT application by replacing
\dComp{s} the application depends on.  Since \DecentComp{s}
execute within TEEs authenticated by RA,
the confidentiality and integrity of the application does not depend
on the trustworthiness of the host or its operator: any host
may provide the service.

We have formalized the \DECENT protocol in ProVerif~\cite{blanchet2016proverif}
and proven it protects the secrecy and authenticity of the data
it processes.
We have also implemented two \DECENT applications
to evaluate the expressiveness and performance of our design.
DecentRide, a decentralized ride-sharing application, and DecentHT, a
distributed hash table.

We evaluated DecentHT's on the YCSB~\cite{cooper2010ycsb} benchmark and compared the
overhead of \DECENT's authorization mechanisms to a traditional
RA approach.
Our results demonstrate that using \DECENT improves
throughput for shorter sessions by as much as 7.5x and latency by as much as 3.67x.
SGX attestation technology such as Intel's DCAP extensions~\cite{scarlata2018dcap},
 and others~\cite{knauth2018ratls,wang2019rustsgx, openenclave} that
avoid interactions with IAS using mechanisms similar to self-attestation
are likely to see similar tradeoffs depending on how often authentication
certificates must be refreshed with IAS.
The source code of \DECENT SDK, DecentRide, and DecentHT including the code for
benchmark have been released on GitHub~\cite{decent-repos}.

The rest of this paper is organized as follows:
§\ref{sec:ExpAndBkgnd} motivates the design of \DECENT using our
decentralized ride-sharing example.
§\ref{sec:SysOverview}
gives a high-level overview and the design of the \DECENT Application Platform.
§\ref{sec:SelfAttest} and §\ref{sec:Authorize}
discuss the details of the \DECENT Authentication and Authorization.
§\ref{sec:Implementation}
outlines our implementation of \DECENT SDK and \DECENT handshake protocol.
Next, §\ref{sec:FormalVrfy} presents the composition and result of the formal
verification for \DECENT.
§\ref{sec:ExampleApp} introduces the two sample applications we implemented with
\DECENT framework.
Moreover, §\ref{sec:Eval}
evaluates the expressiveness and performance of our example \DECENT applications.
Furthermore, §\ref{sec:RelatedWorks}
provides discussions on existing works that are related to \DECENT.
§\ref{sec:Discussion} discusses enhancing data sealing, the extension of \AuthList
for stateless open service support, and our future work of automatic code
verification.
Finally, §\ref{sec:Conclusion}
concludes.

\section{Motivation: DecentRide} \label{sec:ExpAndBkgnd}

To motivate the design of the \DECENT framework, we will use DecentRide, our
decentralized ridesharing application, as a running example.
Ridesharing services match riders to drivers
who pick up one or more passengers and drive them to their desired
locations.

Current ridesharing applications are highly centralized.  All
aspects of the system are controlled by the ridesharing company:
setting prices, suggesting routes, matching riders and drivers, and
processing payments.  Moreover, all the data associated with these
tasks is accessible to the ridesharing companies, raising concerns for
passengers who may wish their travel patterns and other personal data
to be kept confidential.  The dominance of current ridesharing
companies also gives drivers few alternatives when prices or policies
are disadvantageous to the drivers' interests.

\begin{figure}
	\centering
	\includegraphics[width=0.97\linewidth]{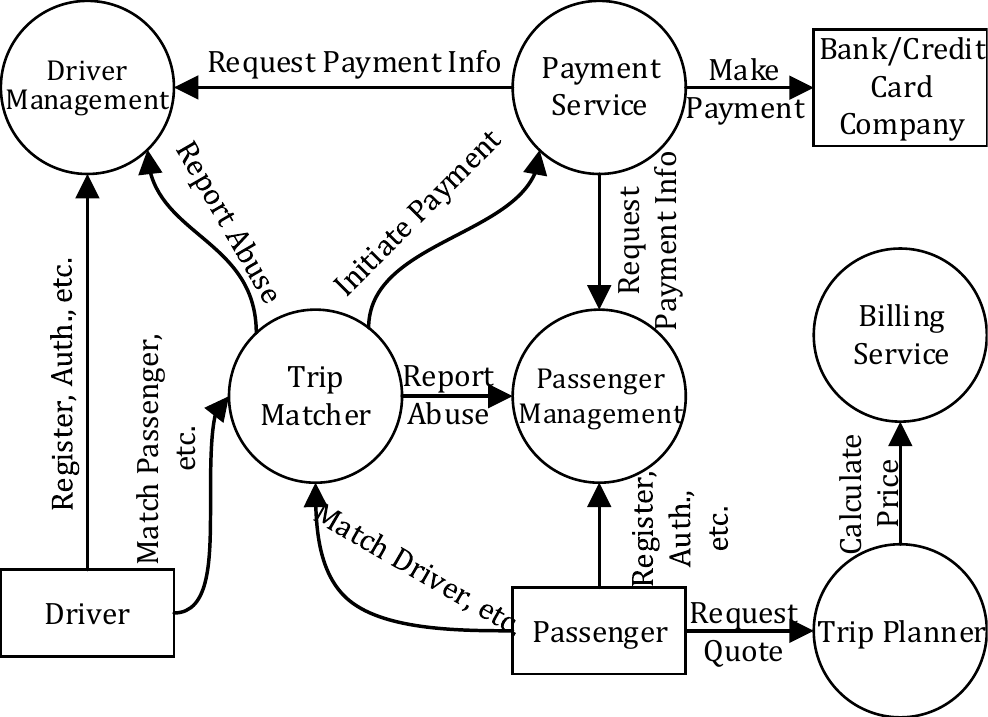}
	\caption{DecentRide, a decentralized ridesharing service.}
	\label{fig:rideshare}
\end{figure}

A decentralized ridesharing application could address some of these
concerns by letting drivers, passengers, and service providers
self-organize, but designing such an application has several security
challenges. Figure~\ref{fig:rideshare} illustrates the
interactions between components of DecentRide, a ridesharing service loosely
based on Uber's microservice-based
design \cite{kappagantula2018microsvc}.

In a decentralized application, these components may be hosted by
multiple entities, some of which may be untrustworthy.  For example, a
malicious entity hosting the Trip Planner component could learn the
locations and routes of drivers and passengers, and a malicious
Billing Service component could manipulate prices.
Trusted execution environments are a useful tool for
implementing DecentRide since remote hosts can establish the
authenticity of a remote component via RA, and
communicate over authenticated, encrypted channels that are
inaccessible to the component's host.  Unfortunately, additional challenges remain.

First, since components may be
hosted by untrustworthy entities, they must mutually authenticate each
other, leading to the circular dependencies described in
§\ref{sec:Intro}.
Resolving these dependencies requires some
authorization data, such as the hash of each component and the operations
allowed for each component, to be provided by the host at load time. It is
critical that this data cannot be used to
subvert the security of data processed by the application.
Furthermore, since some DecentRide components do not
directly connect to each other, they cannot be directly authenticated
and authorized by all components. Therefore, avoiding transitive trust
attacks as illustrated in Figure~\ref{fig:trans} is also critical for
security.  For example, since the Billing Service only interacts with
the Trip Planner, the host of the Trip Planner might attempt
to introduce a malicious Billing Service to
manipulate prices.

Second, distinguishing legitimate updates
from malicious ones is challenging in a decentralized application.
Components that do not exist at compile time, the specifications are created,
cannot be authorized based on the code they contain.
Therefore, an additional mechanism must be used to
authorize new components. The specific authorization process may be
application specific, but any runtime process that authorizes new code
(or revokes the authorization of existing code) should itself be
authorized by the entities of the system whose security is at stake; otherwise,
the process might be used to introduce malicious components.

Finally, because of the high cost of
authenticating components using RA, most TEE
applications establish an ephemeral session key during authentication,
amortizing the cost over the lifetime of the session.  Unfortunately,
since each component in microservice-based designs like DecentRide's
may be replicated in response to demand (and stopped when demand
drops), the lifetime of each component may be relatively short.  These
shorter lifetimes make RA a significant performance
bottleneck, especially for Intel SGX TEEs using EPID-based RA, where one must contact the
Intel Attestation Service (IAS)~\cite{anati2013epid} to determine whether an
attestation is authentic.

\section{\DECENT System Overview} \label{sec:SysOverview}

\begin{figure}
	\centering
	\includegraphics[width=1.0\linewidth]{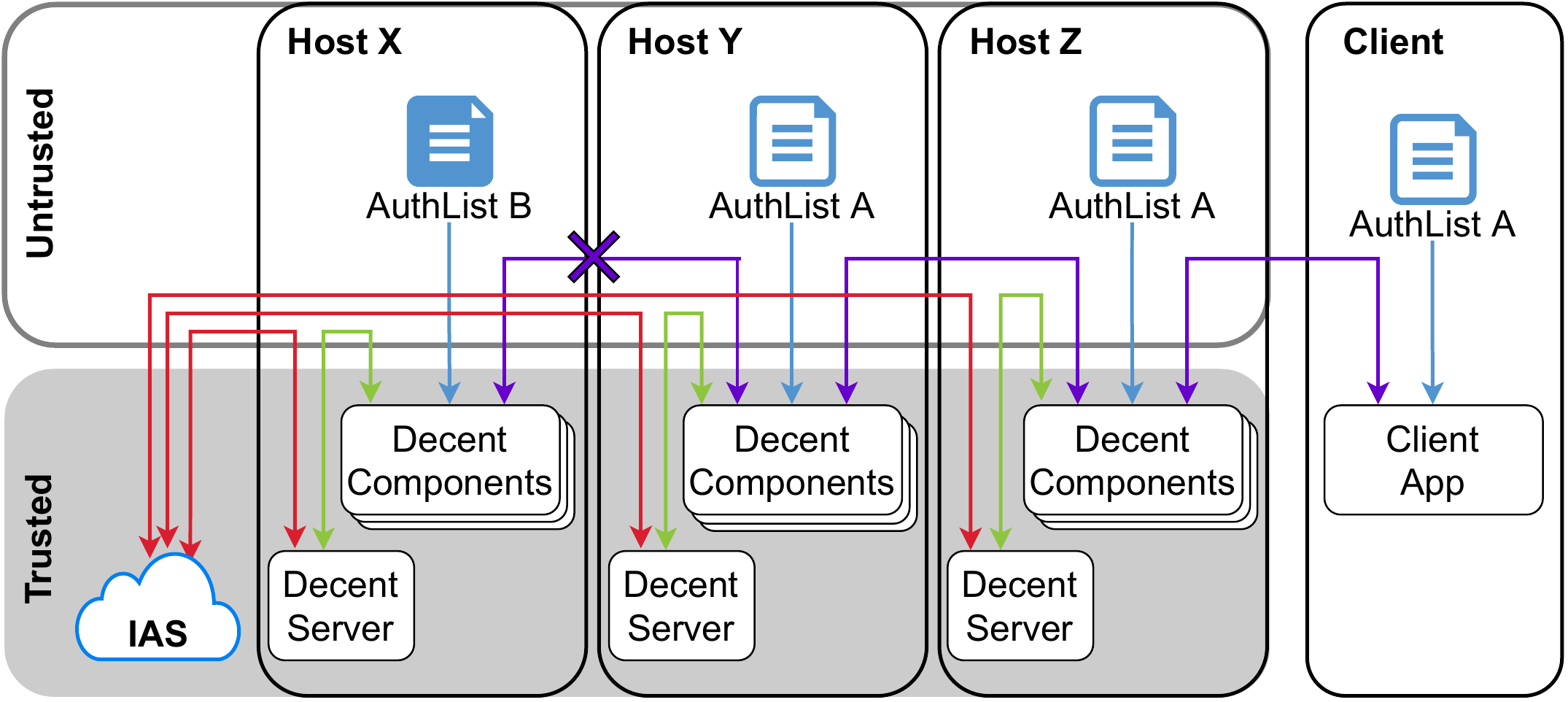}
	\caption{Overview of the \DECENT Framework; trust model shown is from the perspective of a client}
	\label{fig:SysOverview}
\end{figure}

Figure~\ref{fig:SysOverview} provides an overview of \DECENT framework.

\subsubsection*{Building Blocks} \label{ssec:Overview-Frame}

The \DECENT Framework includes \DecentComp{s} and \DecentSvr{s}.
Trusted code in each \DecentComp and \DecentSvr runs in an SGX enclave.
Each host runs a single \DecentSvr and one or more \DecentComp{s}.
The \DecentSvr has only one purpose: to perform self-attestation and
issue certificates to local \DecentComp{s}.
\DecentComp{s} are enclaves that hold \DECENT certificates issued by the
\DecentSvr and follow \DECENT protocols to verify peer certificates.
There are three kinds of \DecentComp{s}: \DecentApp{s}, \DecentVrfy{s},
and \DecentRevc{s}.
\DecentApp{s} contain all functionalities of the application.
\DecentVrfy{s} and \DecentRevc{s}, discussed in
§\ref{ssec:Verifier} and §\ref{ssec:Revc},
concern dynamic authorization and revocation.

\subsubsection*{Authentication} \label{ssec:Overview-Authen}

\DECENT Framework authenticates \DecentComp{s} by using the self-attestation
certificate and \DecentComp certificate.

The first time a \DecentSvr is executed, it creates a key pair
and initiates the RA to bind its public key
to the code running inside the \DecentSvr.
We call this process ``Self-Attestation (SA)'' because the component itself
plays the role of the remote verifier in the protocol.
The goal of SA is not to authenticate the \dSvr to itself
but to create a certificate verifiable by a third party.

SA benefits applications like DecentRide by reducing the
latency overhead of \dComp authentication.
Rather than performing RA directly during the authentication process,
\DecentComp{s} authenticate each other using SA certificates
created at load time.
The traditional approach of establishing an ephemeral key during RA scales
poorly for applications like DecentRide, where \dComp-to-\dComp sessions may
be short-lived.
Because SA certificates are reusable across
sessions, even applications with short sessions scale well in \DECENT.

After \DecentSvr has created its SA certificate,
the \DecentComp sends its \AuthList and its own public key to the \dSvr,
using a secure channel established by the \emph{Local Attestation (LA)},
which provides similar guarantees to RA but does not require verification
by the IAS since the \dComp{s} and the \dSvr reside on the same CPU.
The \dSvr then returns its SA certificate and a signed \dComp certificate
containing the \dComp's public key, the digest of the \dComp's code,
and the \dComp's \AuthList.

Therefore, to authenticate a remote \DecentComp, the verifier needs the
\DecentComp certificate, signed by a \DecentSvr's public key, as
well as the \DecentSvr's SA certificate, so that the
authenticity of the \DecentSvr's key can be verified.
For brevity, we will use \emph{SA certificate} or just \emph{certificate}
as a shorthand for these credentials necessary to verify a \DecentComp.

Factoring out the SA code keeps the SA protocol easier to understand
and audit, and reduces the size of \dComp{s}.
Furthermore, sharing \DecentSvr for \dComp{s} residing on the same CPU reduces
the authentication overhead on hosts that run multiple \dComp{s},
since only one SA certificate is needed per host.

Since each \dComp is running in a separate enclave, they
cannot access the memory of the \DecentSvr or any other \DecentComp, so
even a malicious \dComp cannot tamper with the authentication
process of any other \dComp.

\subsubsection*{Authorization} \label{ssec:Overview-Auth}

To authorize different \dComp{s} into the system,
\DECENT requires all hosts and clients to provide an \AuthList containing
a list of \DecentComp{s} they trust.
They compose the list freely on their own, but only the \DecentComp{s}
and clients who hold exactly the same \AuthList can talk to each other.
That is because the \AuthList held by one \dComp may contain components that the
other one does not trust, and vice-versa.

\DECENT authorization is decentralized in the sense that hosts may include
arbitrary entities for the \AuthList of \DecentComp{s} they host.
However, the contents of the \AuthList constrains which remote \dComp{s} will
accept their connections.
If a group of hosts colluded to include a malicious \dComp, any client or
legitimate \dComp attempting to connect would see the \AuthList is different
and refuse the connection.
As long as clients only connect to \dComp{s} having the same \AuthList, any
\dComp outside of the client's \AuthList cannot communicate with the system
that the client is connecting to.
For instance, as shown in Figure~\ref{fig:SysOverview},
the \dComp running in Host Y received a
different \AuthList from the \dComp in Host X, so the \dComp in Host Y refuses the
connection from the \dComp in Host X.

\subsubsection*{Threat model} \label{ssec:AdvModel}

Figure~\ref{fig:SysOverview} also shows the trust model of \DECENT framework
from the perspective of a client.
Unlike traditional applications, where everything in the figure will be trusted,
\DECENT applications only trust the code running in the TEE environment and
IAS, which is the manufacture of the TEE environment in general.
Thus, the host may provide malicious inputs to \DecentComp{s}
while observe outputs from them.
We assume clients trust their machines, but
clients that support enclaves could potentially execute the client software
in enclaves to mitigate malware attacks.
\DECENT does not prevent vulnerabilities in
code but provides a mechanism for entities in a decentralized
application to agree upon which \dComp{s} should be part of the TCB,
and to verify that only authorized \dComp{s} receive access to
protected data.

We assume that honest \DecentComp{s} follow the \DECENT API specifications
and only communicate with peers that have been authenticated using the
\DECENT protocol.  Honest clients only communicate with \dComp{s}
whose \AuthList consists of \DecentComp{s} that the client considers
trustworthy.  We assume the security of the TEE mechanism: computation
within the TEE is confidential, hosts can only interact with the TEE
via the interfaces defined by the developer, and cannot alter the behavior of
code within the TEE.  We also assume that attackers cannot
compromise cryptographic mechanisms such as public-key encryption or digital
signatures with non-negligible probability.

Given these assumptions, the \DECENT system protects against
adversaries that attempt to subvert security in several ways.  Any
number of hosts in a \DECENT application instance may be malicious.
Malicious hosts may attempt to access and manipulate all inputs and
outputs to \DecentComp{s} and \DecentSvr{s}, including local memory and storage,
messages between \dComp{s}, and configuration data such as \AuthList{s}.
Attackers may also develop and execute malicious \DecentComp{s} and \DecentSvr{s}
that run inside or outside the TEE and partially or fully
violate the \DECENT APIs and protocols.

We do not consider information an adversary learns by analyzing the
timing or pattern of communications outside of the TEE.  Complementary
approaches exist for eliminating leaks from indirect or implicit flows
(e.g., information flow control mechanisms \cite{jfabric,LIO}),
timing channels (e.g., predictive mitigation~\cite{azm10}), and access
pattern analysis (e.g., oblivious
computing~\cite{oblivm,zahur2015obliv}).

The \DECENT platform currently only provides \emph{confidentiality}
and \emph{integrity} guarantees.  TEEs alone
cannot provide \emph{availability} guarantees since untrustworthy
hosts can always suppress messages sent to or from the TEE, or simply
shutdown the TEE altogether.
In §\ref{ssec:BlockRevoke}, we discuss our ongoing work to integrate enclave
with blockchain to achieve confidentiality, integrity, and availability.

\section{Authenticating with Self-Attestations} \label{sec:SelfAttest}

\begin{figure}
	\centering
	\includegraphics[width=1.0\linewidth]{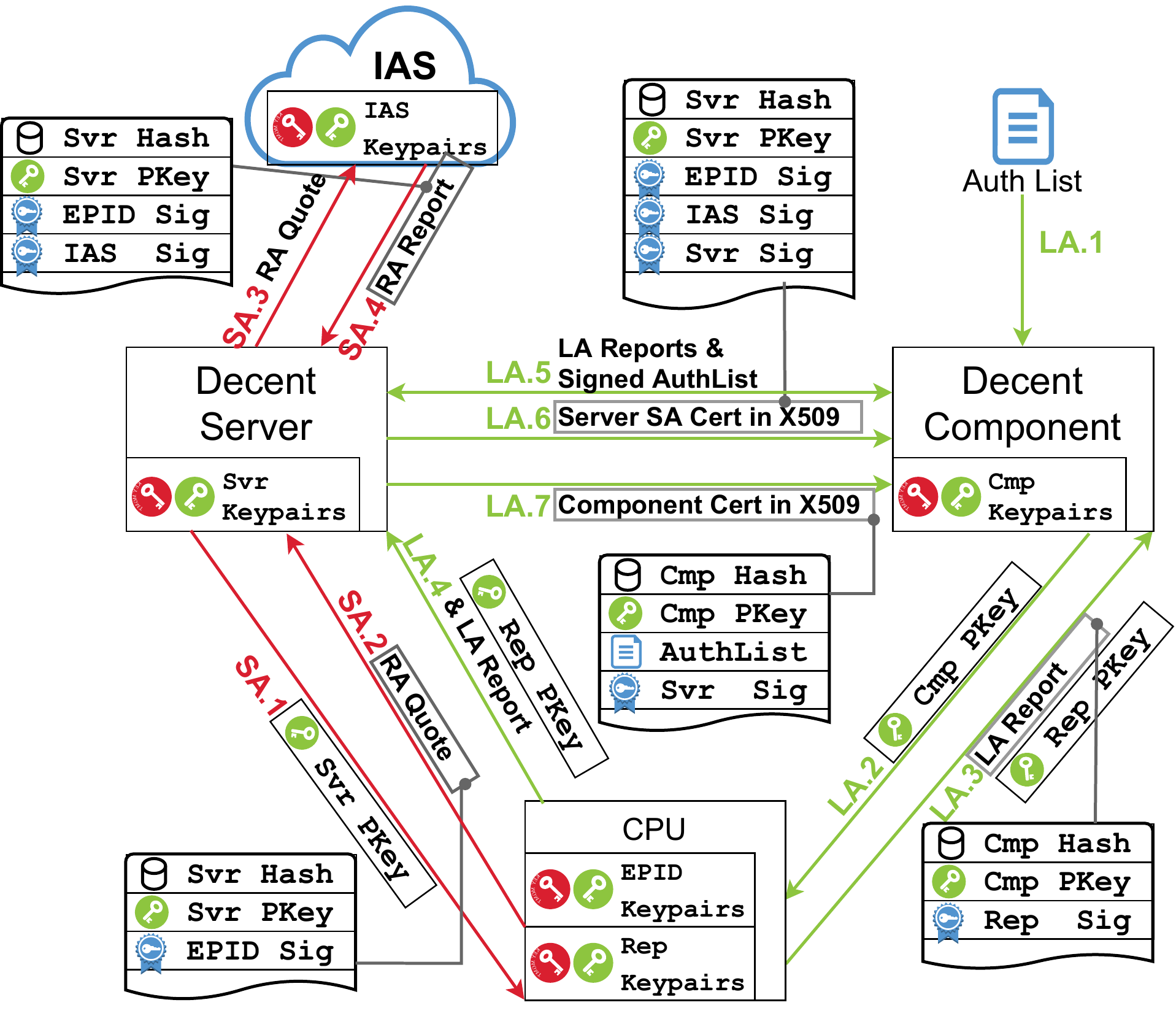}
	\caption{Process of creating certificates for the \DecentComp authentication}
	\label{fig:SelfAttestation}
\end{figure}

Figure~\ref{fig:SelfAttestation} illustrates the process of creating
certificates for the \DecentComp Authentication, which is done in two major
steps: 1) \DecentSvr{s} creating Self-Attestation certificates;
2) Local Attestation of \DecentComp{s}.

In SA process step SA1, the \DecentSvr first creates its
key pair for authentication and sends the fingerprint of the public key to
the CPU to produce a RA quote signed by the EPID~\cite{epid} group signing
key provisioned by Intel to the CPU.
In step SA3, The host forwards the signed quote, returned in step SA2,
to the IAS for verification.
If the signature is valid, in step SA4, the IAS returns a signed RA report,
verifiable by a well-known public key.
Upon receiving and verifying the IAS report, the \dSvr
creates a signed X.509 certificate including the RA report and its public key.

In LA process step LA1, the \DecentComp first loads an \AuthList,
which is immutable once loaded.
It also creates its key pair for authentication and sends the public key
fingerprint to the CPU for signing in step LA2.
Next, in step LA3 and SA4, both the \dSvr and the \dComp request
the CPU to produce a LA report which is verifiable to any enclave running on
the same enclave platform, including the \dSvr , with CPU's public report key.
After verifying each other's LA report, a secure channel is established.
Through this channel, the \dComp sends its \AuthList in step LA5.
Then, in step LA6 and LA7, the \dSvr issues a
\DecentComp certificate containing \dComp's hash and \AuthList signed by the
\dSvr's authentication key, along with the \dSvr's SA certificate.

After \DecentComp{s} have received certificates from the \DecentSvr , they can
communicate over TLS connections that are authenticated with their certificates.
The detailed procedures during the handshake process is given
in §\ref{ssec:DecentHS}.

By verifying the RA report using Intel's public key, third parties can confirm
that the public key in the \dSvr's certificate was created by a specific
\DecentSvr in an authentic SGX enclave.
By verifying the signature on \dComp's certificate, third parties can confirm
the \DecentComp resides in the same enclave platform as the \dSvr , and the
authenticity of the \AuthList.

SA certificates can also be used to verify outputs of a
\DECENT application.  For example, the DecentRide Payment service could
provide digital receipts that verify a user's payment for a particular
trip.  By signing the receipt and attaching its SA
certificate, any third party can verify the contents of the receipt
was produced by a legitimate instance of the DecentRide app containing
no unauthorized \dComp{s}, even if the instance who signs the receipt is no
longer running.

It is also possible to verify attestations without contacting IAS using Intel's
recently added DCAP~\cite{scarlata2018dcap} features (although it is still necessary to
contact IAS for timely revocations).
DCAP allows developers to use a local customized service to verify quotes,
reducing the latency of obtaining quote verification reports from IAS.
Extending \DECENT to support DCAP would be relatively simple:
instead of using the IAS report as the root of the SA certificate,
the DCAP report would be used instead.
The primary requirement for Decent
is that the authenticity of custom DCAP report is verifiable by a third party.

\section{Authorization} \label{sec:Authorize}

	\subsection{Authorization List (\AuthList)} \label{ssec:AuthList}

\begin{figure}
	\centering
    \begin{tabular}{l|c}
      Code Digest & Service Name   \\ \hline
      dff1...8e41 &  BillingService \\
      3fb5...cc46    & PaymentService \\
      6233...0f6d  & TripMatcher  \\
      717e...5c1b  & TripMatcher \\
      ...  &
    \end{tabular}
	\caption{Example \AuthList for DecentRide}
	\label{fig:SysDes-AuthList}
\end{figure}

Figure~\ref{fig:SysDes-AuthList} illustrates an example \AuthList for the
DecentRide example.
Each entry in the \AuthList maps the hash of a
\DecentComp's code to the service name it is authorized to implement.
A service may be provided by multiple enclave binaries to support multiple
platforms and backwards compatibility.

Each \DecentComp creates a unique key pair that is not only
bound (by the SA certificate) to a specific, authentic
instance of an Intel SGX enclave, but is also bound to a specific
\emph{immutable} \AuthList which contains a list of service names and the
\dComp{s} that are authorized to provide them.  That means hosts
cannot modify the \AuthList without launching a new instance of the
\DecentComp.  Malicious \dComp{s} may present a false \AuthList,
but they cannot forge the SA certificate of an
authorized \dComp. Since any authentic attestation report includes
a hash of the component's code, malicious components cannot represent
themselves as authorized ones.

The (untrusted) host establishes the initial network connection and
forwards messages for \DecentComp{s}. The \DecentComp establishes a secure
communication channel using TLS on top of this connection.  \DComp{s}
authenticate themselves by submitting a SA certificate,
and a remote connection from a \dComp is only authorized if the
signatures of all certificates (including the IAS report) are valid,
and the \AuthList{s} match.

	\subsection{Dynamic \DComp Authorization} \label{ssec:Verifier}

Requiring all \AuthList{s} in an application instance to match
prevents unauthorized \DecentComp{s} from connecting to the instance, but
it also prevents new \dComp{s} from being authorized dynamically.
Applications may wish to dynamically authorize \dComp{s} for a number
of reasons, but a common reason is to update \dComp{s} to add
features, improve performance, or fix bugs.  Since \DecentComp{s} are
authorized based on a digest of their code, these new \dComp{s} will
not be allowed to connect to existing application instances.  To authorize
new \dComp{s}, all existing \dComp{s} must be restarted with a new \AuthList.

Requiring full system restarts for \dComp updates goes against the
typical microservice-based design workflow where components are
frequently and independently updated, and multiple versions of a
component may co-exist at runtime.  To avoid the downtime associated
with such restarts, \DECENT distinguishes two special roles that enable
dynamic authorization and revocation: \emph{\DecentVrfy{s}} and \emph{\DecentRevc{s}}.

A \DecentVrfy is also an enclave program,
which is permitted to authorize new \DecentComp{s} (including other \dVrfy{s})
by an application instance.  Comparing to trusted third parties,
trusting a \dVrfy is expressing trust only in the specific code running in the
enclave, and the enclave platform ensures that its behavior cannot deviate
from that code. In addition, when some \dVrfy needs to be authorized by
another \dVrfy, the developers must specify the service names for each level
explicitly, so there is a fixed depth for each valid chain.

\begin{figure}
  \centering
  \includegraphics[width=0.85\linewidth]{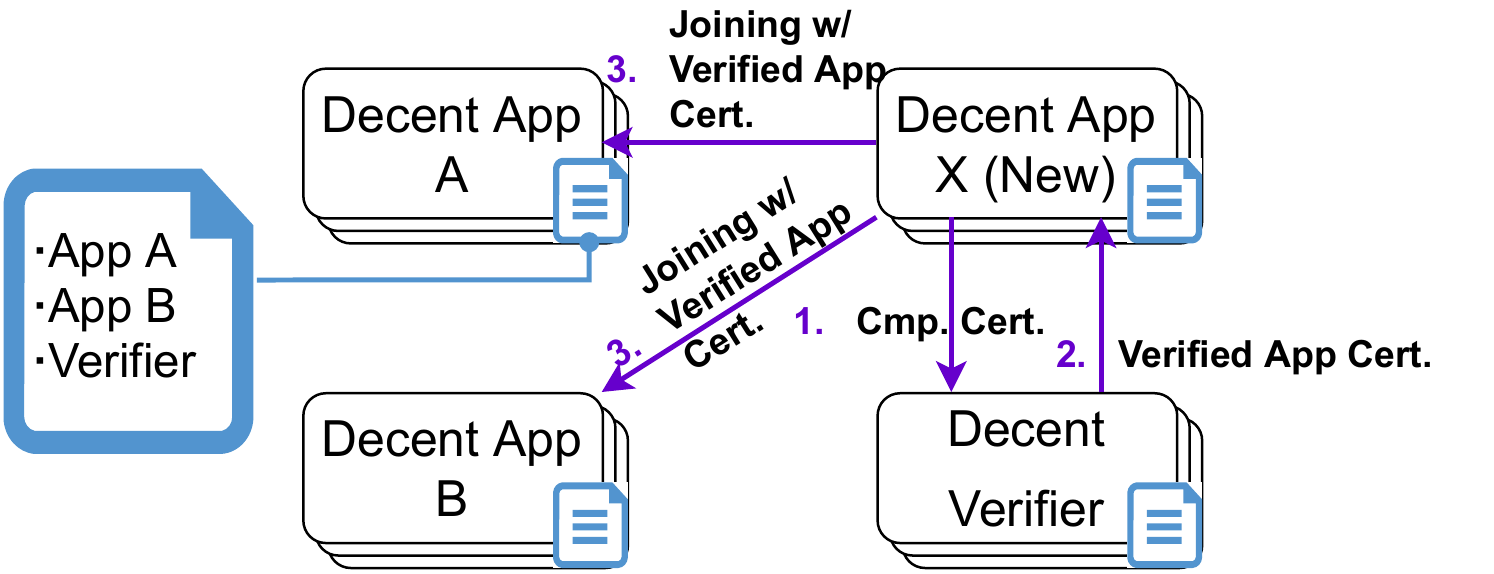}
  \caption{Procedures for the \dVrfy}
  \label{fig:Verifier}
\end{figure}

Figure~\ref{fig:Verifier} provides an overview of \DecentVrfy.
Like any other \DecentComp, in order to join an application instance,
the \DecentVrfy must be listed as a \dVrfy on the \AuthList{s} of the
\dComp{s} in the instance.
In other words, all \DecentComp{s} must agree on which
\dVrfy{s} are authorized to make dynamic authorization decisions.
By defining multiple \dVrfy service names, applications can designate
which \dVrfy{s} are authorized for which services.
For instance, the DecentRide application could define a BillingServiceVerifier
name for authorizing BillingService \dComp{s},
and a TripMatcherVerifier for authorizing TripMatcher \dComp{s}.

The new \dComp first presents its SA certificate to \dVrfy.
\DVrfy{s} process new \dComp{s}' SA certificate through its
verification mechanism defined by developers.
If the certificate is successfully verified, \dVrfy{s} authorize new \dComp{s}
by signing their SA certificate.
Finally, the new \dComp can communicate with the existing \dComp{s} by
presenting its SA certificate signed by the \dVrfy.
A \dComp is authorized to connect to an application instance if (a)
it appears in the \AuthList for the expected service or (b) its
SA certificate is signed by a \dVrfy who appears in
the \AuthList for the expected service \dVrfy.
In both cases the
\AuthList of the new \dComp must match the application instance's.
In the latter case, the \dVrfy's SA certificate must
also contain a matching \AuthList.

Developers may plug in any desired verification mechanism into the \DecentVrfy.
One example approach is for the \dVrfy to collect
signed approvals from an application's ``stakeholders,'' e.g., the
entities whose data security may be affected by the authorization.
When new \DecentApp is created, stakeholders that wish to authorize new \dApp
create and sign new \dApp's code digest.
New \DecentApp authenticates itself to the \dVrfy using its SA certificate.
If the \dVrfy has received the necessary approvals,
it responds by signing new \dApp's certificate.

Other dynamic authorization approaches could
avoid the need for external entities to explicitly authorize new components.
For example, in Fabric~\cite{jfabric} nodes download mobile code
from untrusted hosts, formally verify their information flow properties,
and link the compiled code into a distributed application at runtime.
A similar approach could be adapted for \DecentVrfy.
Each new \DecentComp's source code would be checked by the \dVrfy against a
formal specification associated with the service it claims to implement.
More details about this approach will be discussed in §\ref{ssec:AutoVerify}.

	\subsection{Revocation} \label{ssec:Revc}

Both the \DecentComp and the SGX platform can be compromised.
During such a event, the private key of the \DecentComp could be leaked, so that
attacker can pretend to be a \DecentComp and join the system,
while the damage is specific to application.
BFT (Byzantine Fault Tolerance) protocol could help application to tolerate
compromised nodes, but it is beyond the scope of this paper.
Regardless, the ability to revoke vulnerable component is key to addressing
such problem once it’s detected.

\DecentComp{s} may have their SA certificates revoked
in one of two ways. First, the IAS maintains a
number of revocation lists it uses to determine
whether an SGX platform (the chip, the credentials provisioned to it,
or the platform software itself) have been revoked.  Since RA
reports are signed by a group signature to protect
privacy, only Intel is able to distinguish revoked platforms within a
group. Even with DCAP enabled, enclaves still need to acquire revocation lists
from IAS
A \DecentComp running on a host whose platform has been revoked
will be unable to refresh its SA certificate with the
IAS server. Therefore it is prudent to set SA
certificates to expire at reasonable intervals to force periodic
refreshing.\footnote{While a malicious host may manipulate the
operating system clock, the SGX platform provides a trusted interval
timer that can be used as the basis of a mechanism to measure
certificate lifetimes and force refreshes.  We leave the design and
implementation of such a mechanism to future work.}
Additionally, compared to the original RA protocol,
saving the RA result to the certificate
and refreshing it periodically does not weaken the security guarantee.
According to the sample code in Intel SGX SDK, to reduce the overhead
caused by RA, it is suggested that the RA state is also kept by
secret provisioning for a period of time.
Thus, the revocation of the a platform will also only be discovered
after the RA state is refreshed.

\subsubsection*{Dynamic \dComp revocation}

If authorized \DecentComp{s} are discovered to have latent
vulnerabilities, or are incompatible with new updates, these
\dComp{s} should no longer be permitted to connect to a \DECENT
application instance.  However, dynamically revoking \dComp
authorizations presents many of the same challenges as dynamic
authorization. Forcing full system restarts to modify \AuthList{s}
leads to increased downtime, and may delay when revocations could reasonably take
effect. Furthermore, modifying \AuthList{s} does not address revocation
for dynamically authorized \dComp{s}.

\DECENT revokes authorized \dComp{s} using Component Revocation Lists
(CoRLs).  A CoRL is similar to a Certificate Revocation
List~\cite{rfc5280}, but
instead of revoking a specific SA certificate, a CoRL
entry is used to revoke \emph{any} SA certificate
generated for a specific \DecentComp.
In other words, that component may no longer connect to the instance,
regardless of who is hosting it.

\DecentComp{s} called \emph{\DecentRevc{s}} maintain the CoRLs associated
with an application instance.  Like \dVrfy{s}, \dRevc{s} must appear on
the \AuthList{s} of all \DecentComp{s} in the application instance, or
must themselves be authorized by a \dVrfy.

\begin{figure}
	\centering
	\includegraphics[width=0.85\linewidth]{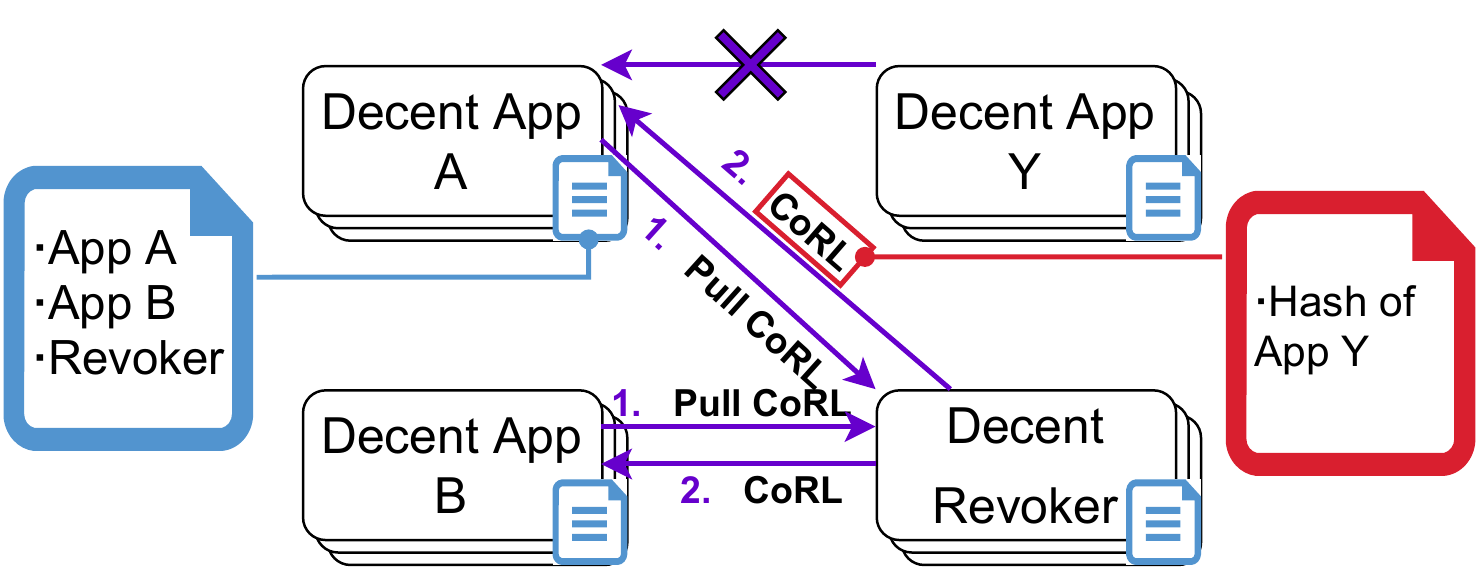}
	\caption{Procedures for the \dRevc}
	\label{fig:Revoker}
\end{figure}

Figure~\ref{fig:Revoker} illustrates a revocation workflow.
The \DecentRevc must also be listed as a \dRevc on the \AuthList{s} of the
\dComp{s} in the instance.
\DComp{s} periodically pulls CoRL from designated \dRevc{s}, to ensure they are
having the latest CoRL locally.
\DComp{s} shuts themselves down if no responds is received from the \dRevc , to
prevent subsequent damage causing by malicious host suppressing revocation
message.
When a revoked \dComp attempts to communicate with other \dComp{s} , it will be
detected when other \dComp{s} consulting their local CoRL.

Similar to \dVrfy , developers can also plug in any revocation mechanism to
the \dRevc.
For example, stakeholders submit revocation requests to one or more
\dRevc containing the hash of the \dComp whose authorization is to be revoked.
Once a threshold of revocation requests are received,
the \dComp's hash is added to the CoRL.

In some scenarios, it may be possible to revoke \dComp{s} without
the intervention of an external entity. If evidence that a \dComp{s}
is compromised is mechanically verifiable, \dRevc{s} can offer API that allows
nodes to submit messages or private keys as evidence of compromise, so that
\dRevc{s} can automatically add entries to the CoRL.
Our \DECENT prototype does not yet support \dRevc{s}.

\DRevc{s} may revoke the authority of any \DecentApp or
\DecentVrfy. Revoking the authority of a \DecentSvr or \DecentRevc
dynamically is problematic. The \DecentSvr is designed to be small
and rarely updated.  Many \dComp{s} in an application instance will
likely share the same \DecentSvr, so revoking the server would
invalidate the \DECENT certificates of many \dComp{s}.
Moreover, \DecentSvr are not allowed to be dynamically authorized, so these
\dComp{s} would be unable to rejoin unless there is a different
version of authorized \DecentSvr that has not been revoked.
Revoking a \dVrfy can similarly cause many \dComp{s}' SA certificates to be
rejected, but unlike the server scenario, these \dComp{s} may rejoin as
long as they are able to locate some other authorized \dVrfy that belongs
to the instance.

\DECENT prohibits the revocation of \DecentRevc{s} for two
reasons. First, the desired effect of revocation is unclear.
Entries on CoRL from a revoked \dRevc might not be replicated on the
CoRLs of other \dRevc{s}, so discarding those entries might allow
compromised \dComp{s} to rejoin the application instance.
Accepting entries from a revoked \dRevc is also inadequate, because some
entries may be erroneous
or malicious (hence the revocation), but we cannot identify which ones are.
Second, if two different \dRevc{s} place each other on their CoRL,
which revocation should take precedence is unclear.
Finally, \dRevc{s} may be dynamically authorized by \dVrfy{s}, but these
\dVrfy{s} must be treated specially. Since revoking the authority of
a \dVrfy of \dRevc{s} would lead to the same issues that accompany revoking
a \dRevc directly, the \dVrfy{s} of \dRevc{s} cannot have their authority
revoked dynamically.  To avoid these issues,
no CoRL is consulted when authenticating \dRevc{s} or their verifiers.

\subsection{Distributed \dRevc{s}} \label{ssec:BlockRevoke}

In the previous section, we introduced a simple revoker setup, where all
\DecentComp{s} periodically poll for updates to the CoRL from a \dRevc, allowing them to
automatically shutdown to prevent further damage if the attacker is trying
to suppress messages from the \dRevc.
A single revoker service creates a point-of-failure for ensuring the authority of vulnerable components can
be effectively revoked.
To reduce load on the \dRevc and tolerate \dRevc failures, system designers could replicate
the revoker service and use a consensus protocol to ensure the CoRL is consistent across
replicas.  In this section we describe a design that distributes CoRLs using a Proof-of-Work (PoW) blockchain
such as Ethereum.

\begin{figure}
	\centering
	\includegraphics[width=1.0\linewidth]{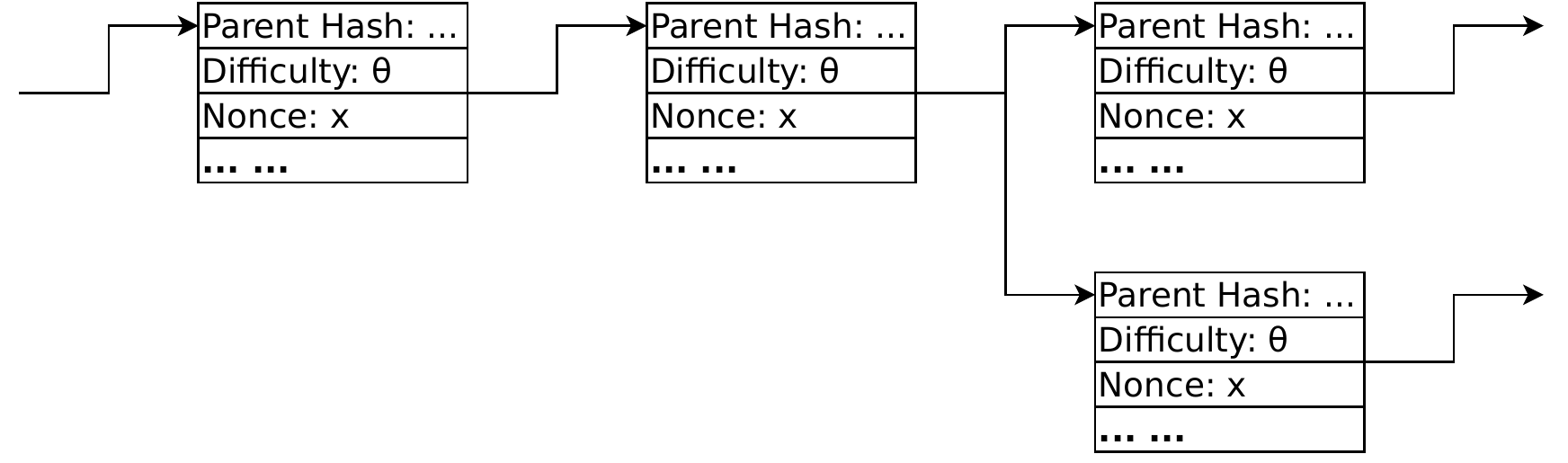}
	\caption{Blockchain chain structure}
	\label{fig:BlockchainChain}
\end{figure}

Figure~\ref{fig:BlockchainChain} shows the general chain structure of a
blockchain.
Each block contains the hash of its parent block, a difficulty value of the
puzzle that miners are going to solve, and the nonce which is the answer to
the puzzle.
When mining a new block, miners work concurrently to find the nonce that
solves the puzzle whose difficulty is specified in the parent block, and
multiple miners may find an answer before the next block is widely distributed.
When miners mine different blocks pointing to the same parent,
a fork (or forks) is formed.
If there are multiple forks, miners choose which fork to extend, picking the longer
chain if one exists.
In order to ensure new blocks are generated at a relatively
steady rate, the difficulty value is adjusted based on the time it took for
a miner to mine and distributed the new block.

For applications with trusted stakeholders available, developers may
choose to use permission-based ledger, where only authorized nodes (e.g., the stakeholders)
may create new blocks.
However, in case that stakeholders are unavailable, a public available
blockchain network like Ethereum is needed.
In Ethereum, any node can follow the protocol to mine new blocks, and there is
and assumption that at least $50\%$ of miners are honest.
Otherwise, attackers could mine the longest chain and have the blocks to only
include the transaction they prefer.
An attacker may still want to maintain a private malicious fork and distribute
it to other peers.
This is not an issue in a regular blockchain client, which can connect to as
many peers as possible and hope that at least one of them is honest, so that
it can always obtain the view of the honest chain.
the honest chain is longer than the malicious chain because of the assumption
that the majority of miners are honest.
However, in enclave applications, all network connections are routed through
the OS.
Thus, it is possible that none of the blockchain message received by the
enclave are coming from the honest node.

Zheng et al. \cite{haofan2021eclipse} have proposed an approach to detect
such an attacker by monitoring the difficulty value in the block.
The attacker only has limited hash power, so they usually take longer time
to mine the new block.
As the block time increases, the difficulty value will drop according to DAA.
Therefore, by monitoring the difficulty value, the enclave can determine if the
blocks are from malicious fork or not.

By storing the CoRL on blockchain, multiple \dRevc instances can
update/retrieve the same CoRL as well as serve multiple \dComp{s}.
Meanwhile, by utilizing the smart contract feature of Ethereum blockchain,
we can even automate the processing of adding new \dComp{s} into the CoRL.
For instance, when the private key of a \dComp is found outside the enclave,
it can be submitted to the smart contract, which will verify the key and add
the corresponding component into the CoRL.
Additionally, if a compromised \dComp is sending conflicting messages, we can
submit the messages to smart contract to add it to the CoRL.

\section{Implementation} \label{sec:Implementation}

We have implemented a prototype of our design using
Intel SGX for Windows.
Our prototype consists of about 20k lines of C++ (12k excluding header files) and
uses Intel SGX SDK version 2.3.101.50222, and Mbed TLS version 2.16.0.
While some of our design decisions are informed
by the constraints of the SGX platform, our high-level design is
applicable to any TEE platform that supports RA and
secure memory.

	\subsection{Using the \DECENT SDK} \label{ssec:DecentSDK}

The \DECENT SDK provides a high-level API for
establishing secure channels between \dComp{s} that greatly
simplifies authentication and authorization of remote application
\dComp{s} while preserving fine-grained control over which \dComp{s}
are authorized to perform specific services.

System calls such as those that handle network
connections cannot be executed within an SGX enclave.
The standard
approach~\cite{knauth2018ratls} for establishing a secure channel with
an enclave is for untrusted code to first create (or accept) a TCP
connection to (or from) the remote host, and then act as a proxy
between the enclave and the network.

\begin{figure}
	\centering
\begin{minted}{c++}
void proc_trip_matcher_req(void* cnt_ptr)
{
	//Get DECENT authorization data:
	//key pair, certificate, auth list, etc.
	States& state = GetStateSingleton();

	EnclaveCntTranslator connection(cnt_ptr);

	//Configure TLS session
	std::shared_ptr<TlsConfigWithName> tlsCfg =
		std::make_shared<TlsConfigWithName>(
			state,
			TlsConfig::Mode::ServerVerifyPeer,
			"TripMatcher",
			GetSessionTicketMgr());

	//Create TLS channel
	TlsCommLayer tls(
		connection, tlsCfg, true, nullptr);
	tls.Recv(/* ... */);
	tls.Send(/* ... */);
}
\end{minted}
	\caption{Creating a TLS channel with \AuthList}
	\label{fig:SampleCode-TlsWithName}
\end{figure}

Figure~\ref{fig:SampleCode-TlsWithName} shows a fragment of code from
the Payment Service \dComp in DecentRide that handles incoming
requests from the Trip Matcher \dComp.  The \texttt{void*} pointer
\texttt{cnt\_ptr} points to a TCP connection created by the untrusted
code and passed into the enclave.  The authorization data of
this instance of \dComp is retrieved on line 5, and a wrapper class for the TCP
connection is instantiated on line 7.  Lines 10-15 create an object
that configures how the connection should be authenticated and
authorized.  The authorization data \texttt{state} is passed in to
provide the TLS library with the \dComp's key pair, certificate
chain, and \AuthList.  The mode \texttt{ServerVerifyPeer} indicates
that the Payment Service is expecting an incoming request (hence
Server) and should request a certificate from the remote \dComp
(hence VerifyPeer). Line 14 specifies that the expected name
of the remote \dComp is "TripMatcher", thus the name "TripMatcher"
must be listed under the \dComp's hash entry in the \AuthList.  On line
15, "GetSessionTicketMgr()" retrieves a reference to a TLS session
ticket manager that helps resume sessions to avoid re-negotiating the
TLS handshake unnecessarily.  Lines 18 and 19 establish the TLS
channel using the connection wrapper and the configuration object, and
lines 20 and 21 use the channel to receive and send data with the
remote \dComp.

Permitting TripMatcher \dComp{s} to be authorized using a \dVrfy
only requires a few modifications to lines 10-15 of the
Payment Service code above:
instead of instantiating the "TlsConfigWithName" object, we create a shared
pointer to a "TlsConfigWithVerifier" object, whose constructor accepts
an additional \dComp name, "TripMatcherVerifier", for the \dVrfy's
name permitted to authorize "TripMatcher" \dComp{s}
dynamically.
This configuration requires that any \dVrfy{s} of a "TripMatcher"
\dComp be listed under the name "TripMatcherVerifier" in the
\AuthList.

\subsection{\DECENT Handshake Protocol} \label{ssec:DecentHS}

\begin{figure}
	\centering
	\includegraphics[width=1.0\linewidth]{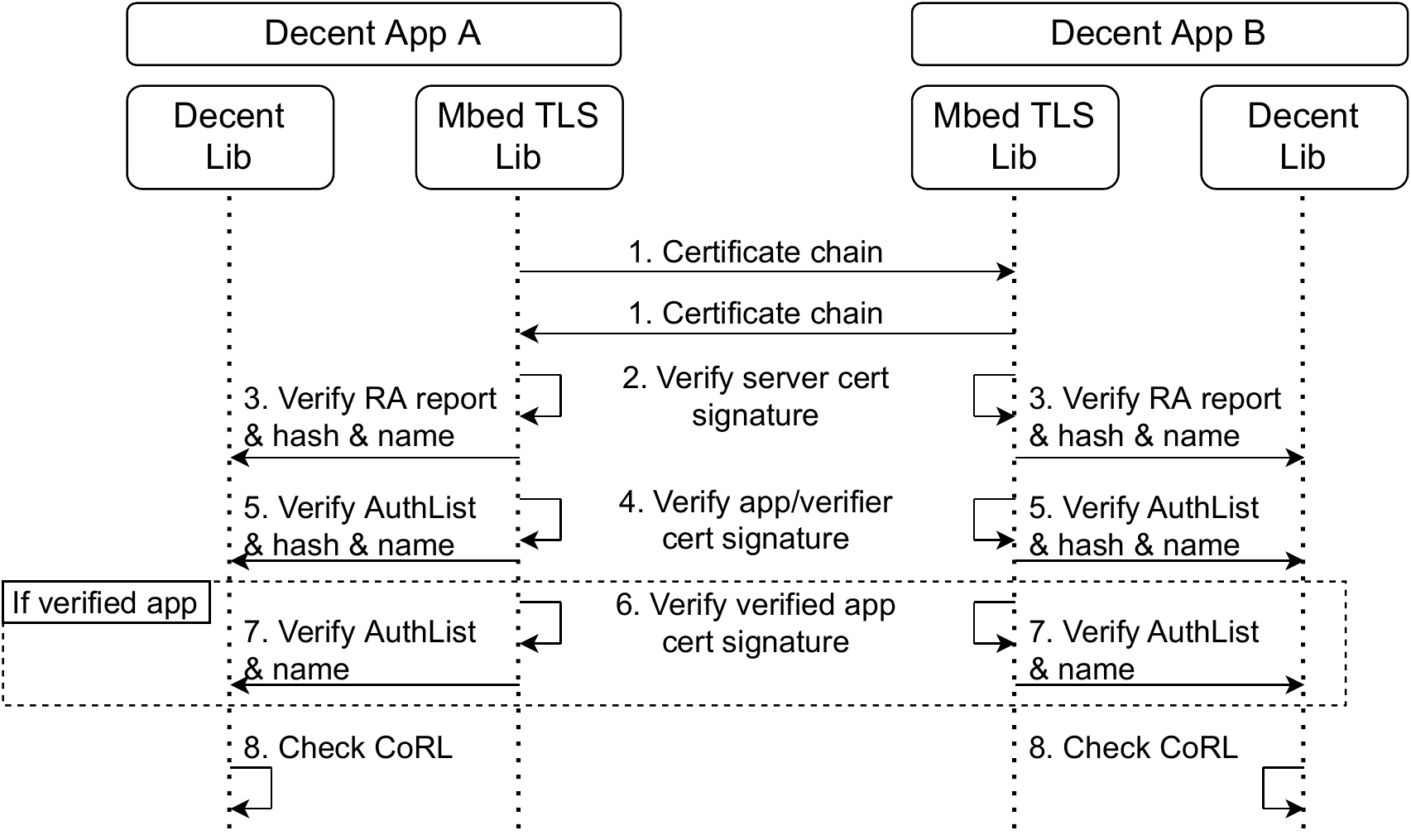}
	\caption{\DECENT handshake workflow}
	\label{fig:Impl-Handshake}
\end{figure}

The \DECENT handshake extends the usual TLS handshake where both sides
authenticate with X.509 certificates.
The \texttt{Mbed TLS} library code will handle most TLS handshake process,
including the certificate exchanging step.
If signatures are successfully verified, a callback function implemented in
\DECENT is executed by \texttt{Mbed TLS} to verify the customized contents in
the X.509 certificate.
Figure~\ref{fig:Impl-Handshake} illustrates the handshake procedure between
two \DecentApp{s}.

\begin{enumerate}
	\item \DecentApp A and B exchange their X.509 certificate chains, which includes:
	\begin{itemize}
		\item The X.509 certificate of \DecentSvr containing the RA report
		\item The X.509 certificate of the \DecentApp or \DecentVrfy signed by \dSvr
		\item The X.509 certificate of the \DecentApp signed by the \DecentVrfy , if
		the \DecentApp is verified by the \dVrfy .
	\end{itemize}

	\item The Mbed TLS library code validates the signature on \DecentSvr's certificate

	\item The RA report contained in \DecentSvr's certificate is verified with
	Intel's public Report Key, and the local \AuthList is consulted to ensure the \DecentSvr's
	hash appears under the ``\texttt{DecentServer}'' service name

	\item The Mbed TLS library code validates the signature on \DecentApp's (or \dVrfy's) certificate

	\item Our callback function will ensure the \AuthList found in the certificate
	matches the one we loaded, and the hash of \DecentApp (or \dVrfy) found in
	the LA report appears under the expected service name in the \AuthList

	\item If the remote peer is a verified app, there will also be a certificate issued
	to the verified \dApp by a \dVrfy; The Mbed TLS library code validates the
	signature on \DECENT verified app's certificate

	\item The callback function will ensures the \AuthList found in the certificate
	extension section matches the one we loaded, and its service name matches the
	one defined in the code

	\item The local revocation list is consulted to ensure all hashes—\dApp{s},
	and \dVrfy{s}—are still valid and have not been revoked
	\footnote{Our current prototype does not implement revocation yet.}
\end{enumerate}

If all checks are successful, the connection is accepted, otherwise it is rejected.

\section{Formal Verification} \label{sec:FormalVrfy}

We have formalized and verified the secrecy and authentication properties of the
\DECENT protocol design using ProVerif~\cite{blanchet2016proverif}.
ProVerif is an automated formal verification tool for
cryptographic protocols,
which we use to formalize our protocol's behavior and
describe the scenario we want to test.
The tool allows us to ask whether the secret could be revealed to the attacker
or if the attacker can manipulate a message without being detected.
The protocol is public to attackers, and attackers may intercept and manipulate
any (raw) message transmitted within message channels.

Our ProVerif formalization consists of 1095 lines of code and comments for the
implementation and 1437 lines for the verification scenarios,
and it follows the threat model defined in §\ref{ssec:AdvModel}.
We use the ProVerif standard library for cryptographic definitions
used in our implementation, which assumes that encryption and digital signatures
are uncompromisable if appropriately used.

\begin{figure}[h]
	\centering
	\includegraphics[width=\linewidth]{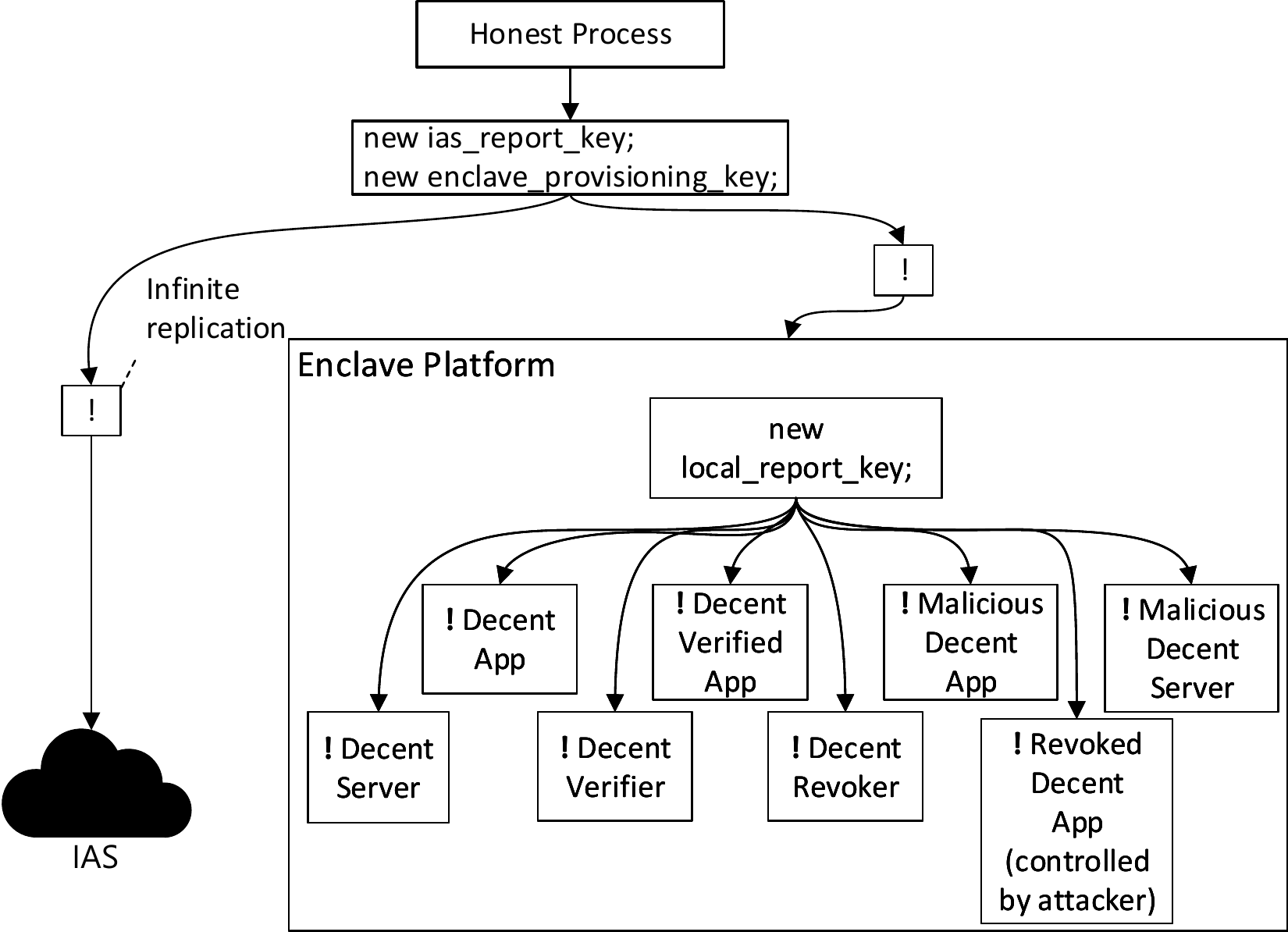}
	\caption{Verification Processes Overview}
	\label{fig:VrfyProc}
\end{figure}

As shown in Figure~\ref{fig:Verifier},
we defined five process types representing \DECENT's architecture building blocks:
\DecentSvr, \DecentApp, \DecentRevc, \DecentVrfy, and
Verified \DecentApp (a \DecentApp authorized by a Decent Verifier).
We also defined
additional processes for modeling malicious \DecentComp{s} and \DecentSvr{s}.
For simplicity, we only consider verification of \dComp{s} from the perspective of
(honest) \DecentApp{s};
the verification process is identical for clients of \DECENT services.

Each \DecentApp process loads an \AuthList at the beginning of the process.
One honest \dApp contains secret data to protect, or will receive a computation result
whose authenticity must be verified.  We designate the \AuthList of this \dApp to be the
``reall'' \AuthList, meaning it contains only trustworthy components.
The \AuthList{s} of all other processes are given by an attacker representing the untrusted host.
The authorized \DecentComp{s} and \DecentSvr{s} are represented by honest processes, since
a host cannot alter their behavior.

Attacker-controlled \DecentComp{s} and \DecentSvr{s} are expressed by issuing RA
and LA reports in an honest process,
but leaking the \dComp's private key to the attacker.
Hence, attackers will be able to authenticate as the \dComp without being bound
by the original process's behavior.
Note that the IAS key, CPU EPID key, and CPU report keys discussed in
Figure~\ref{fig:SelfAttestation} remain secret.
We assume revocation lists and hashes of approved \DecentApp{s} are
provided to \DecentRevc{s} and \DecentVrfy{s} via out of band process.

We verified the secrecy and authenticity of the data transmitted between
\DecentComp{s} using two verification scenarios.
In one scenario, there are two \DecentApp{s}: one sending the
secret data, the other receiving the data.
The other scenario is identical, but between Verified \DecentApp{s}.
Any of these processes may be replicated an arbitrary number of times.
Figure~\ref{fig:VrfyProc} illustrates
the process diagram for our formalization.

\begin{figure}
\begin{tabular}{l | l | l}
                     & \DecentApp   & Verified \DecentApp \\
  \hline
   Secrecy        &  2 minutes       & 8 hours                  \\
   Authenticity  & 4 hours           & 61 hours
\end{tabular}
\caption{Verification times}
\label{tbl:proverif}
\end{figure}

Table~\ref{tbl:proverif} lists the verification times for our ProVerif formalization.
Verifying data secrecy for \DecentApp{s} completed relatively quickly,
which only took 2 minutes,
while verifying secrecy for Verified \DecentApp{s} required 8 hours.
We also verified the authenticity of \DecentApp{s},
which took 4 hours to complete.
Verifying authenticity for Verified Apps required decomposing the verification
problem into three simpler tasks.
First, we prove that the \AuthList stored in the certificate issued by
\DecentVrfy{s} is identical to the \AuthList loaded by the Verified \DecentApp.
Next, the \dVrfy only issue certificates to Verified \DecentApp{s} with the
same \AuthList loaded.
Lastly, we prove that Verified \DecentApp{s} only accept peers with identical
\AuthList{s} in their certificates.
The entire process took 61 hours to complete.
Our complete verification code and reports are available on GitHub~\cite{decent-repos}.

\section{Example \DECENT Applications} \label{sec:ExampleApp}

	\subsection{DecentRide}

We used our prototype to implement two \DECENT applications.  The first
application is an implementation of DecentRide, discussed in
§\ref{sec:ExpAndBkgnd}.  DecentRide \dComp{s} are hosted by a
distributed network of mutually-distrustful nodes.
Because these
components run inside SGX enclaves, confidential information like the
location and routes of drivers and passengers is kept private, and
critical computations like route and pricing calculation cannot be
manipulated by the hosts.

Each DecentRide \dComp (circles in Figure~\ref{fig:rideshare})
implements a microservice—a small, targeted
service that is loosely coupled with the other \dComp{s} by a simple
application protocol. With the exception of the TripMatcher,
DecentRide \dComp{s} maintain no local state. This design makes it
easy to launch (or halt) \dComp replicas on demand and balance
request load among hosts.  DecentRide hosts are incentivized to
host \dComp{s} by compensation they receive (calculated by the Billing Service)
from payments for trips (processed by the Payment Service).
Passengers and drivers register and authenticate to the Passenger Management
and Driver Manager services, and receive credentials
that are presented to (and verified by) the DecentRide \dComp{s}
they interact with.  All user data provided to the management services
is authenticated encrypted with keys derived
from a \dComp's seal key and the \AuthList as described in §\ref{ssec:AuthList}.
Thus, hosts have no access to the user data they store, even if they launch malicious
DecentRide instances.

When a passenger wishes to find a ride on DecentRide, they contact a
Trip Planner service which finds a path between their current location
and the desired destination. The Trip Planner submits the proposed
route to the Billing Service to get a total price for the trip,
including fees for the hosts operating DecentRide \dComp{s}.  This
price quote is signed by the Billing Service and returned to the
passenger (via the Trip Planner) along with the Billing Service's
certificate chain, which the passenger and Trip Matcher uses to verify the authenticity
of the price quote.

Drivers advertise their availability by sending their location data to a Trip Matcher,
which replies with a list of nearby passengers and their desired destinations.
Once a driver selects a passenger, their contact information will be exchanged,
and the passenger will receive the current location of the driver.
At the destination, both driver and passenger must confirm to the Trip Matcher
that the trip completed successfully, and only then does the Trip Matcher
initiate payment for the trip with the Payment Service.  The Payment Service
obtains payment information from the management services and
submits the transaction to the relevant financial entities.

To prevent abuse such as fake trip requests designed to probe for
driver locations, or fake driver locations to probe for passenger
locations and destinations, driver and passenger interactions with the
Trip Matcher are logged to their respective management services.  The
Trip Matcher will not proceed with a driver or passenger request until
it receives confirmation from the service that the logged action was
successfully received, ensuring that malicious users cannot bypass the abuse
detection by suppressing these logging messages.

	\subsection{DecentHT} \label{ssec:Dht}

To complement to DecentRide,
we implemented a simple Distributed Hash Table (DHT) in
\DECENT based on the Chord~\cite{chord} protocol.
Running each node
within an enclave lets us store confidential information at untrusted
hosts and control access to that information using \DECENT \AuthList{s}.
Sealed data stored in the DecentHT can be accessed based on a consistent hash
function applied to the desired key, just as in the Chord system.
A non-enclave alternative to DecentHT could store encrypted
\emph{values} that were inaccessible to their hosts, but
controlling access to the decryption keys introduces extra complexity in such
a decentralized system.

DecentHT nodes encrypt their data using their own sealing keys and
only provide access to authorized entities. It requires no additional
key management beyond the \DECENT authentication mechanisms.
Executing the Chord protocol logic within the enclave rules out
attacks based on manipulating protocol messages or routing attacks
since even malicious hosts process messages with trusted code.
The host may still, of course, suppress incoming or outgoing messages,
and may still learn some information from analyzing
communication patterns between nodes, but at a much slower rate when
there is a high ratio of keys to nodes. Moreover, an additional data replication
scheme is needed since neither enclave nor \DECENT guarantees availability, and
some versions of DecentHT nodes could be revoked at any time.

DecentHT derives each node's identifier, which determines
the items it is responsible for, using the
\DECENT seal key.  This approach prevents many Sybil attacks~\cite{sybil} since
every DecentHT \dComp launched with the same \AuthList on the same
CPU will receive the same identifier.  \DComp{s} launched with different
\AuthList{s} will be unable to connect to other DecentHT instances that have
agreed on the same \AuthList.

\begin{figure}
		\centering
		\includegraphics[width=0.7\linewidth]{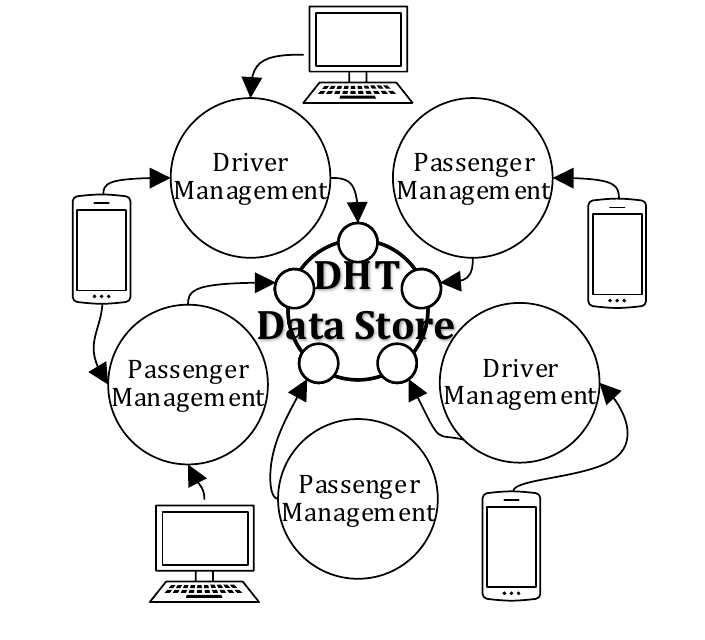}
		\caption{DecentHT: a Decent implementation of Chord~\cite{chord}.}
		\label{fig:decentht}
\end{figure}

Figure~\ref{fig:decentht} illustrates a DecentHT network used by a
DecentRide instance to store passenger and driver data used by the
management services.  Drivers and passengers interact with the
management services, which may cause them to create, fetch, or modify
data stored in DecentHT.
As in Chord, requests will be routed to the appropriate node.
When the \dComp responsible for the key is
reached, the \dComp loads the associated (sealed) data into the enclave
and sends it to the requesting application.

\section{Performance Evaluation} \label{sec:Eval}

	\subsection{Experiment setup}

We evaluate the overhead of \DECENT authentication and authorization by
comparing the performance of DecentHT on the YCSB
benchmark~\cite{cooper2010ycsb} to an implementation that uses a
RA approach as suggested by the Intel SGX SDK
documentation.
Since the SGX RA Only approach does not support mutual authentication of the DecentHT
\dComp{s}, we omit code for enclave identity verification.
In our results, we refer to the \DECENT implementation as \DECENT RA and
the SGX SDK implementation as SGX RA Only.

We also evaluated the performance of two
non-enclave implementations to examine the baseline performance
of the system without the overhead of SGX operations.
In TLS Only group, the DecentHT code executes outside of the enclave
and communicates over TLS channels without verifying certificates.
In TLS+Sealing group, we additionally encrypt
the stored records with a symmetric key, similar to how the enclave versions
seal the records.%

Our experimental setup is similar to management services accessing data stored
in DecentHT, as discussed in §\ref{ssec:Dht}; but
instead of DecentRide \dComp{s}, we created a small \DecentApp
that exposes Java bindings for the YCSB benchmark to invoke.
We used Workload B containing $95\%$ reads and $5\%$ writes,
with uniform request distributions for all our experiments below.
Each read/write operation consists of one lookup request to determine
the node responsible for storing the desired record, followed by a request to
read/write the record.
Each DecentHT node loads approximately 3,000 records.
To ensure a relatively even distribution of records across
nodes, we disabled DecentHT's Sybil resistance and explicitly
specified node ID's that were evenly spaced.
Each test measures
performance for 60 seconds after a 60-second warm up phase.
We repeat each test \emph{three times} and report the median of measurements as
points on the graph, with minimum and maximum values as error bars, which
are \emph{not distinguishable} at the scale of the evaluation graphs.

DecentHT nodes execute on a single 3.6 GHz Intel i7-7700 with 4 cores (8
logical) running Windows 10.  The server has 16 GB of RAM, with
128MB (the maximum permitted) dedicated to SGX.
Records stored at each node are sealed and stored in non-enclave memory.
Each node is assigned its own logical core, with two cores reserved for the
network stack and the Intel AESM service which is responsible for managing
interactions between the operating system and SGX enclaves.

Clients execute on a single 3.4 GHz Intel i3-7100T with 2 cores (4
logical) running Windows 10.  The client machine has 4GB of RAM and
128 MB is dedicated to SGX.
The client and server machines are connected via 1-Gigabit Ethernet.

SGX requires that all thread-local memory be pre-allocated, thus the maximum
number of threads used by an SGX application is fixed at runtime and
is limited by the memory available to SGX. For the DecentHT nodes, we
specified 18 threads for handling incoming requests from either clients or peers,
6 threads for forwarding the finger table lookup requests to other peers,
and 2 threads for replying requests received from other peers.
In the SGX RA Only experiments, 14 additional threads are used for requesting
quotes for the enclave itself.
These threads are unnecessary for the \DECENT
experiments, but we reserve the same amount of memory in both cases.
Using the additional memory to allocate more threads for the \DECENT experiment
would further improve \DECENT's performance over the SGX RA Only case.

On the client side, we limited YCSB to a maximum of 50 threads due to the
enclave memory required by each thread.
This limitation did not affect the throughput of the SGX-based
implementations, which were fully loaded at around 40 client threads.

To test the performance of DecentHT in different session lengths, we run the
experiment with various numbers of requests per session;
the more requests in each session, the lengthier the session is.
The DecentHT nodes perform a full TLS handshake (or RA in SGX RA only group)
in each session's first request,
and the following requests resume the session with TLS session
tickets~\cite{rfc5077}.
The RA process results in a shared secret between the
enclave and the host verifying the enclave's attestation report.
Thus, In SGX RA only group, we used a similar approach to the sample code
provided by the Intel SGX SDK to establish a secure
channel using AES-GCM encryption between the DHT and the application nodes.
In addition, we give nonces and randomized Initial Vectors (IVs)
to the encryption process for all messages, and implemented a
mechanism similar to TLS session tickets~\cite{rfc5077}.
The purpose of this protocol is simply to approximate
the security guarantees of \DECENT's authenticated channels using
secrets shared during the attestation process and avoid the additional
overhead of a TLS handshake, for a fair comparison.

Since our experiments generate many IAS requests in the
SGX RA Only case, we use a simple IAS simulator to avoid violating the
terms of use for our Intel Developer account.
Instead of sending requests to the IAS, all requests are sent to our simulator
which replays a single hardcoded response from the official IAS.
Requestors follow the same protocol as they would for a real IAS request,
but they ignore the nonce in our simulated response to allow us to replay the
same response multiple times.
The response times of our simulated
IAS are \emph{gamma-distributed} with parameters estimated from
\emph{30 IAS API response time samples} collected from the IAS
portal.  For retrieving the current EPID signature revocation list, we
measured a mean response time of 39 ms with standard deviation 24 ms.
For retrieving an attestation report we measured a mean
response time of 255 ms with standard deviation 70 ms.

	\subsection{Results} \label{ssec:ExpRes}

\begin{figure}
	\centering
	\includegraphics[width=1.0\linewidth]{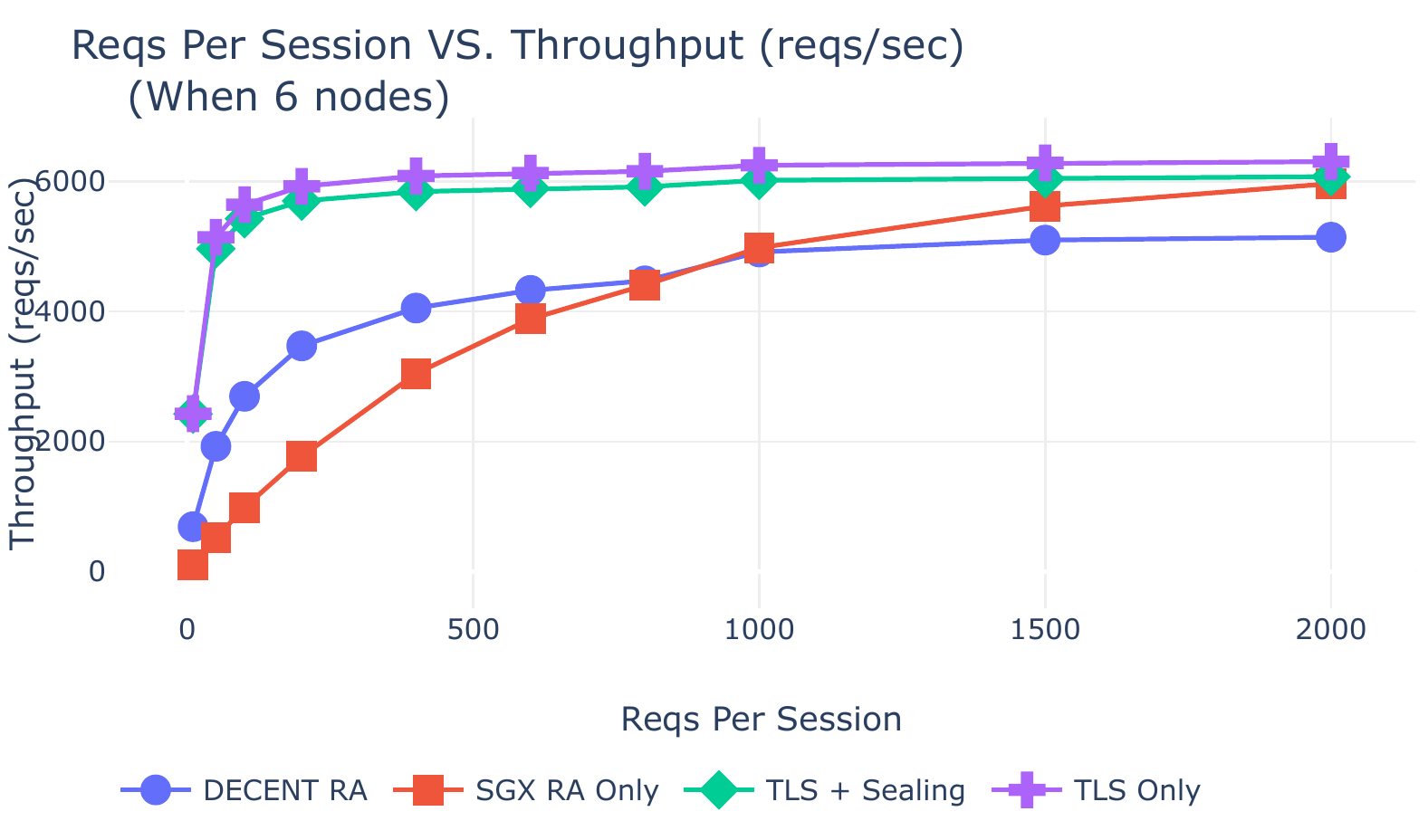}
	\caption{Requests per session versus throughput}
	\label{fig:Eval:OpPerSesVsThr}
\end{figure}

Both the \DECENT RA implementation and the SGX Only implementation
amortize the cost of authentication over the length of a
session. Therefore the negative impact of authentication on system
throughput will decrease as the average session length increases.  To
analyze the tradeoff between authentication overhead and session
length, we ran each implementation with six server nodes on the YCSB
benchmark while varying the number of requests each client made before
establishing a new session.  For the \DECENT RA implementation, each
new session involved a TLS handshake and the exchange and verification
of SA certificate chains.  For the SGX RA Only
implementation, each new session required a new RA.
The non-enclave versions required only a TLS handshake with no
certificate verification.

Figure~\ref{fig:Eval:OpPerSesVsThr} presents the results of these
experiments.  The non-enclave performance improvement plateaus at
about 200 requests per session, with the TLS+Sealing implementation
achieving a lower throughput due to the extra
overhead of encrypting and decrypting the stored records.
 Between 10 and 400 requests per session, \DECENT RA
significantly outperforms the SGX RA Only implementation.
 For long sessions of 800 requests, \DECENT RA and SGX get roughly the same
throughput. Beyond 800 requests, the SGX RA Only performance
approaches the TLS Only implementations, which we attribute to
the additional messages required to resume TLS sessions compared to
our custom SGX RA session ticket scheme.

\begin{figure*}[t]
	\centering

	\begin{subfigure}[t]{0.333\linewidth}
		\centering
		\includegraphics[width=\linewidth]{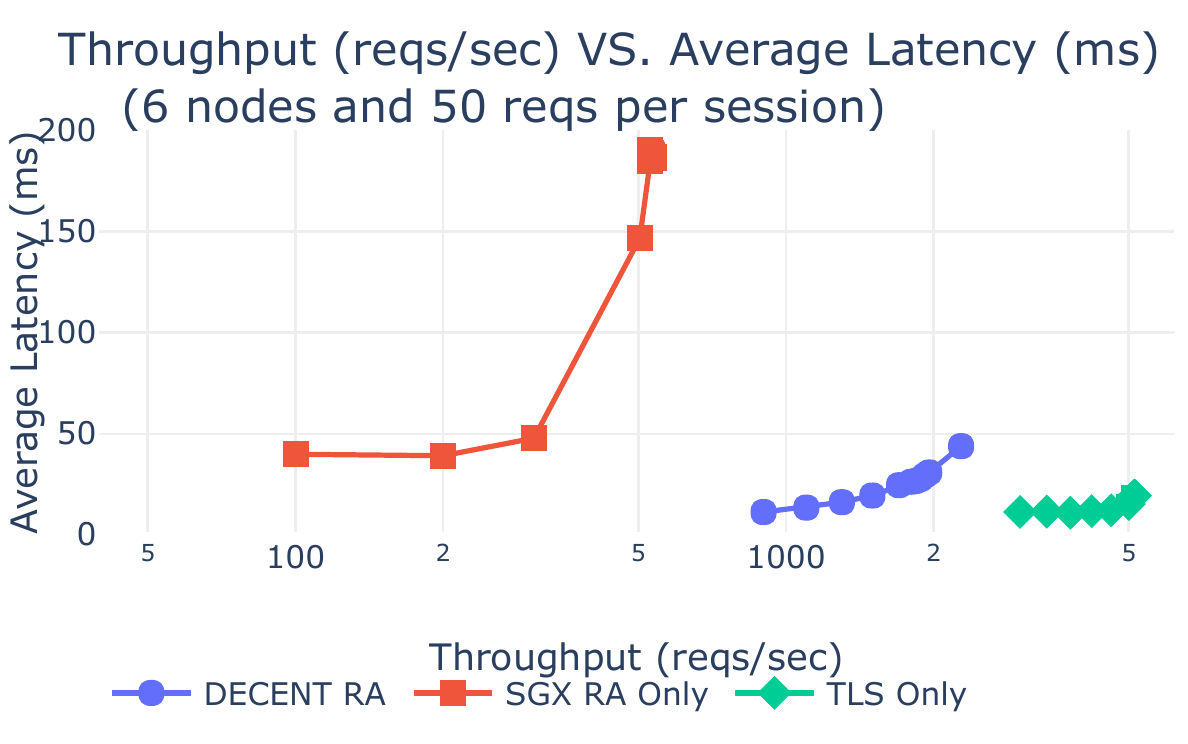}
		\caption{Each session includes 50 requests}
		\label{sfig:Eval:AvgLatVsThr50reqs}
	\end{subfigure}%
	~
	\begin{subfigure}[t]{0.333\linewidth}
		\centering
		\includegraphics[width=\linewidth]{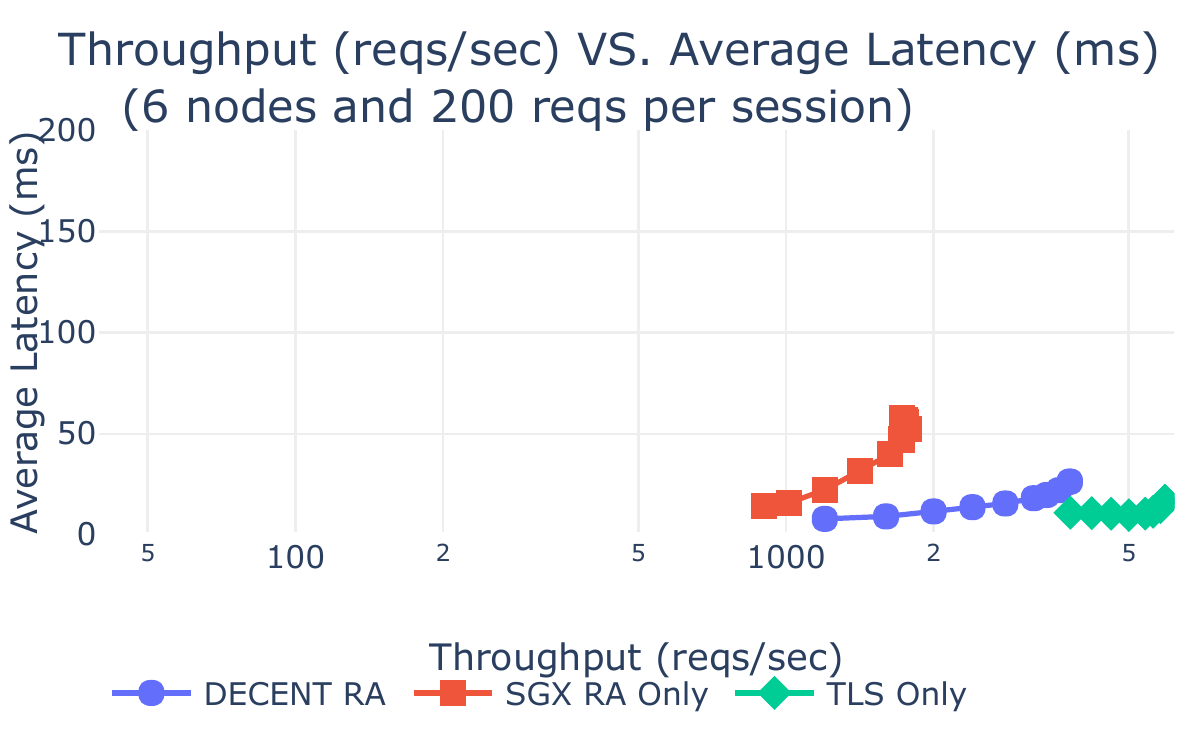}
		\caption{Each session includes 200 requests}
		\label{sfig:Eval:AvgLatVsThr200reqs}
	\end{subfigure}%
	~
	\begin{subfigure}[t]{0.333\linewidth}
		\centering
		\includegraphics[width=\linewidth]{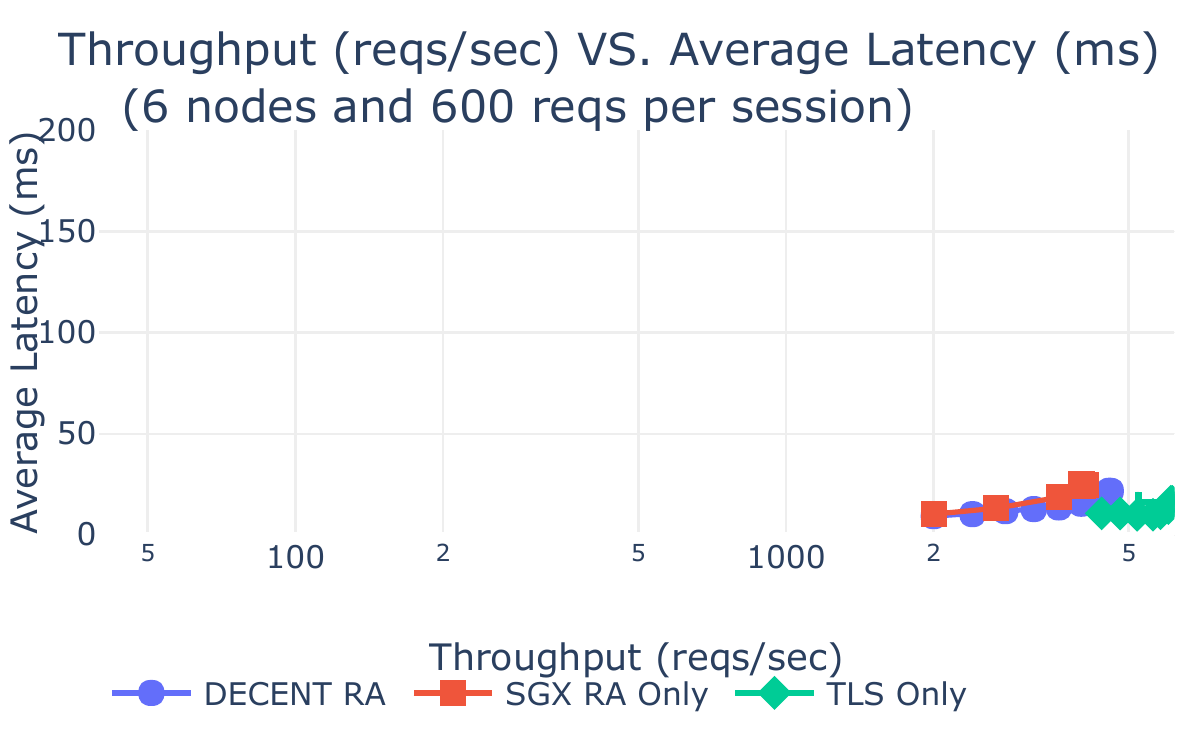}
		\caption{Each session includes 600 requests}
		\label{sfig:Eval:AvgLatVsThr600reqs}
	\end{subfigure}

	\caption{Average latency versus throughput}
	\label{fig:Eval:AvgLatVsThr}
\end{figure*}

Figure~\ref{fig:Eval:AvgLatVsThr} plots the tradeoff between latency
and throughput for sessions of length 50, 200, and 600. We gradually
increased the target throughput and measured the average response time
for requests.  At 50 requests per session, the difference between
\DECENT RA and SGX RA is most pronounced, with latency rapidly
increasing for SGX RA as throughput exceeds 150 requests per second.
At 200 request per session, the behavior begins to converge, but
\DECENT RA still significantly outperforms SGX RA.  At 600 requests per
session, however, their performance is roughly equivalent.  Note that
the performance of the TLS Only implementation is mostly unaffected
for these session lengths.

\begin{figure}
	\centering
	\includegraphics[width=1.0\linewidth]{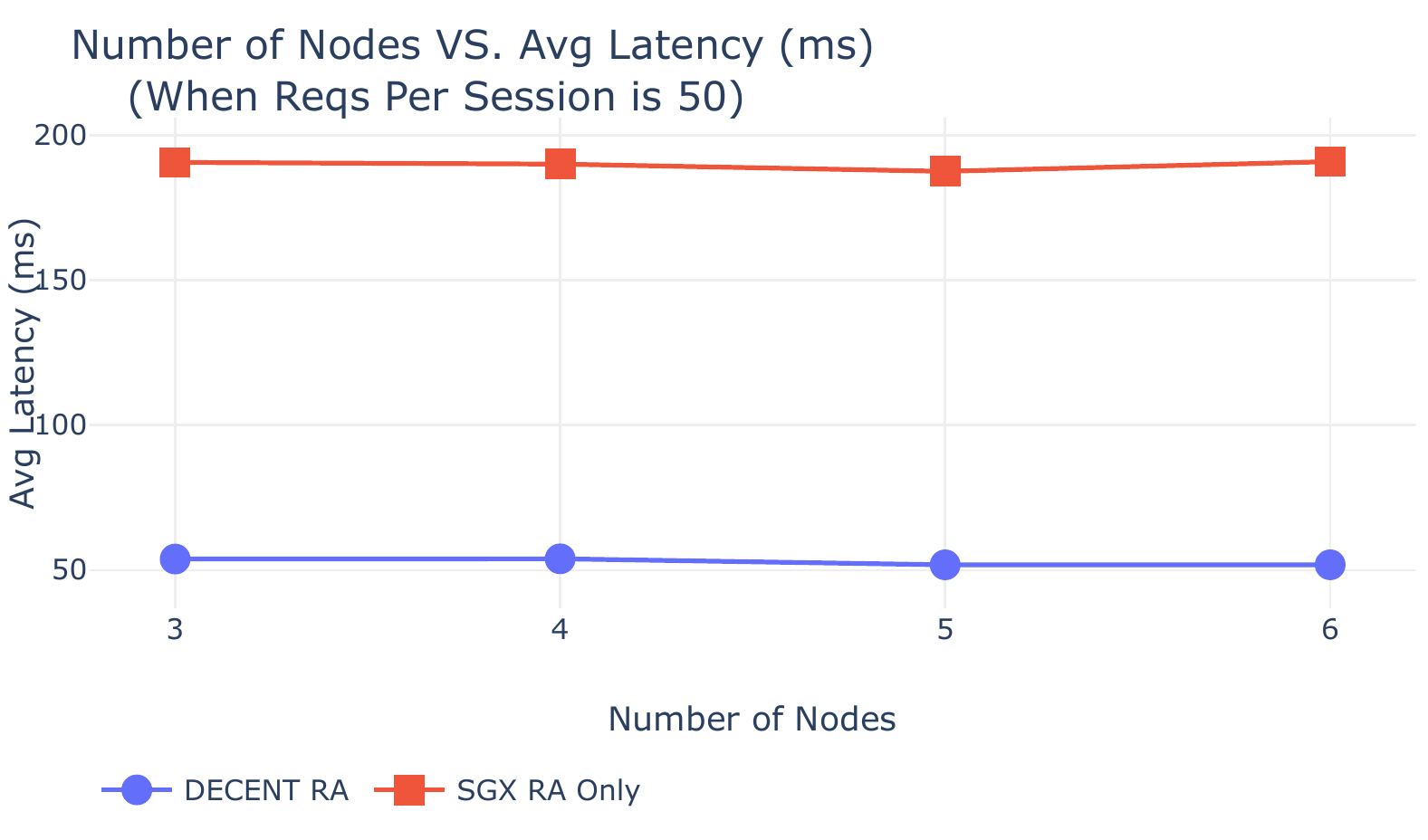}
	\caption{Average Latency vs. Number of Nodes}
	\label{fig:Eval:AvgLatVsNumNode}
\end{figure}

We also measured the latency of requests as the number of nodes increases from
three to six to sanity-check whether \DECENT RA affects scalability.
In Figure \ref{fig:Eval:AvgLatVsNumNode}, the results show that, as expected,
latency in both implementations is largely unaffected by the small increase in
nodes.
However, the average latency for \DECENT RA is almost four times
lower than SGX RA.

\section{Related Work} \label{sec:RelatedWorks}

Several recent projects use enclaves and RA to provide
confidentiality and/or integrity for distributed applications, but almost none
address the problem of mutual authentication. Beekman et al.\cite{beekman16at}
suggest a work-around for an application with two components by building both of
them into one enclave binary and using a runtime switch to select the desired component.
This approach is impractical in a distributed application with large number of components,
since it results in (potentially very) large enclave binaries and is difficult to maintain.
This approach also requires rebuilding the entire binary when any binary requires updating.
VC3~\cite{schuster15vc3} allows a user to launch MapReduce workloads using
cloud-hosted SGX enclaves, ensuring
the confidentiality of the processed data and the integrity of the
results.  VC3 jobs distribute a single enclave binary to each host, avoiding
mutual authentication issues.

Other systems deal with mutual authentication using a external party.
Ryoan~\cite{hunt2018ryoan} authenticates the components
using RA, but modules
are identified by a public key that signs the module rather than the code itself, while
the corresponding private key could be a key that is stored outside of an enclave.\footnote{The authors
claim that Ryoan could also support identities based on code hashes, but it is unclear how they would
address mutual authentication.}
MesaTEE~\cite{sun2019mesatee}
solves the problem of mutual authentication by relying on
third-party \emph{auditors}~\cite{sun2019mutualauth}, which sign
binaries that pass an audit process.  \DECENT
does not require trust in external entities (which could be compromised) to authorize enclaves, but requires
hosts in an application instance to agree upon which components
are authorized to implement which services, and enables clients to
verify which components are included in an instance before using its services.
Panoply~\cite{shinde2017panoply} partitions applications into multiple
small enclave components.
It relies on a shim library to assign names to each enclave and maintain the mapping from name to enclaves'
hash. However, since mutual authentication is not its main focus,
details regarding how the shim library obtains the hash and enforces the mapping make it difficult
to assess how or whether it supports mutual authentication similar to \DECENT's.

CCF \cite{russinovich2019ccf} uses a distributed ledger to manage which TEE components
are enabled. This fills a similar role to the \AuthList in \DECENT.  CCF
nodes are more heavyweight than \DECENT nodes because they must participate in a
consensus protocol to process requests. Since CCF relies on consensus protocols
to protect the integrity of the list of authorized components, the security of CCF against
attacks by malicious components relies on the security assumptions of these protocols (e.g.,
$>2/3$ of hosts are honest, for BFT protocols). \DECENT offers stronger integrity:
regardless of how many hosts are honest, no malicious components may be introduced into
an application instance.  \DECENT does not offer availability guarantees, but coupling \DECENT
with a BFT protocol could provide availability guarantees similar to CCF without sacrificing integrity.

\DECENT's SA process is similar to a now-common approach
to authenticating TLS connections with enclaves.  Knauth {\it et
al.}~\cite{knauth2018ratls} describe the process of using RA to bind a public
key to an enclave that generated it and
include the report in a self-signed certificate.  This certificate is
used to establish the TLS connection allowing the remote host authenticate the key.
Rust SGX~\cite{wang2019rustsgx} and Open
Enclave~\cite{openenclave} offer similar support. Similarly, OPERA \cite{chen2019opera}
proposes an RA service that is separated from the SGX protocol, but
still requires a report generated by IAS during its preparation phase.

All the above TLS approaches support, in principle, the option of
SA where the host of the enclave participates on both
sides of the attestation process to aid in the creation of the
certificate. None address the issue of mutual authentication
of enclaves and therefore cannot support applications like DecentRide
or DecentHT.  Furthermore, our results in §\ref{sec:Eval}
quantify the performance gains of SA over on-demand
attestation for short sessions.

Other work has focused on supporting more general systems programming
within SGX enclaves.  SCONE \cite{arnautov2016scone} and
Graphene-SGX~\cite{tsai2017graphene} support standard library functions not
natively available to enclaves, such as filesystem and network I/O.
In general, these projects focus more on the local secure systems
programming aspects of using SGX enclaves, and do not directly support
RA.

Intel's DCAP support permits customized SGX RA protocols
without contacting the IAS (except during setup). Alternative TEE designs such as Sanctum~\cite{costan16sanctum} and
Keystone~\cite{lee2019keystone} allow direct verification of a public key certificate chain signed by
the manufacturer. Keystone also supports additional roots of trust beside the manufacturer.
DCAP, Sanctum, and Keystone's approaches to RA could reduce the
overhead of RA similar to using SA
certificates. They do not however, address mutual attestation between enclaves.

Key Separation and Sharing (KSS) is a recent feature added to the SGX
SDK that help differentiate multiple instances of the same
enclave. For example, \DECENT could use KSS to derive different seal
keys for each application instance by placing a hash of the \AuthList
in the configuration parameters instead of using our HKDF approach
(as discussed in §\ref{ssec:sealed}).
Furthermore, since we can specify an \AuthList in terms of each
\dComp's \texttt{MRENCLAVE} identity, and bind it to the
configuration id used for attestation, KSS could provide an alternate
implementation path for \DECENT's authorization mechanism.
Nevertheless, although KSS provides additional tools for mutual authentication,
it doesn't provide a solution.  Any KSS-based approach would need to address the same
challenges as \DECENT, but like DCAP, KSS could provide an alternate implementation
path for some of \DECENT's features.

\section{Extensions and Future Research Directions} \label{sec:Discussion}

\subsection{Binding Data Sealing Keys to \AuthList{s}} \label{ssec:sealed}

Enclaves use cryptographic algorithm to protect code and data stored in memory.
However, the space of memory is limited comparing to permanent drives, and
data stored in memory is volatile.
Thus, in many cases, the enclave application also needs to write data to
permanent drives by using the process called data sealing:
the enclave derives the data seal key from its root key and encrypts the data
with authenticated encryption algorithm using the derived seal key.
Intel SGX provides two key-derivation schemes for deriving \emph{seal keys}.
In one scheme, \texttt{MRSIGNER}, the enclave platform uses the
\emph{signer} (typically the author) of the enclave to derive seal
keys, meaning that any enclave binary signed by the same key may decrypt
sealed data.
In the other scheme, \texttt{MRENCLAVE}, the enclave platform uses the enclave's
code hash to derive keys, thus, when multiple enclave applications writing
sealed data to the drive, they can only decrypt the data that has been sealed
with their own key.
\texttt{MRENCLAVE} scheme prevents one enclave application from reading the data
that has been sealed by another one.
The \texttt{MRENCLAVE} scheme is useful for \DECENT applications
since we do not want \dComp{s} exchanging data outside of mutually-attested
channels or malicious \dComp{s} reading data sealed by an honest \dComp.
However, since this approach derives identical keys for each instance of an enclave,
a host could use a malicious \AuthList to attempt to unseal data encrypted by an
application with a secure \AuthList.

\begin{figure}
	\centering
	\includegraphics[width=0.9\linewidth]{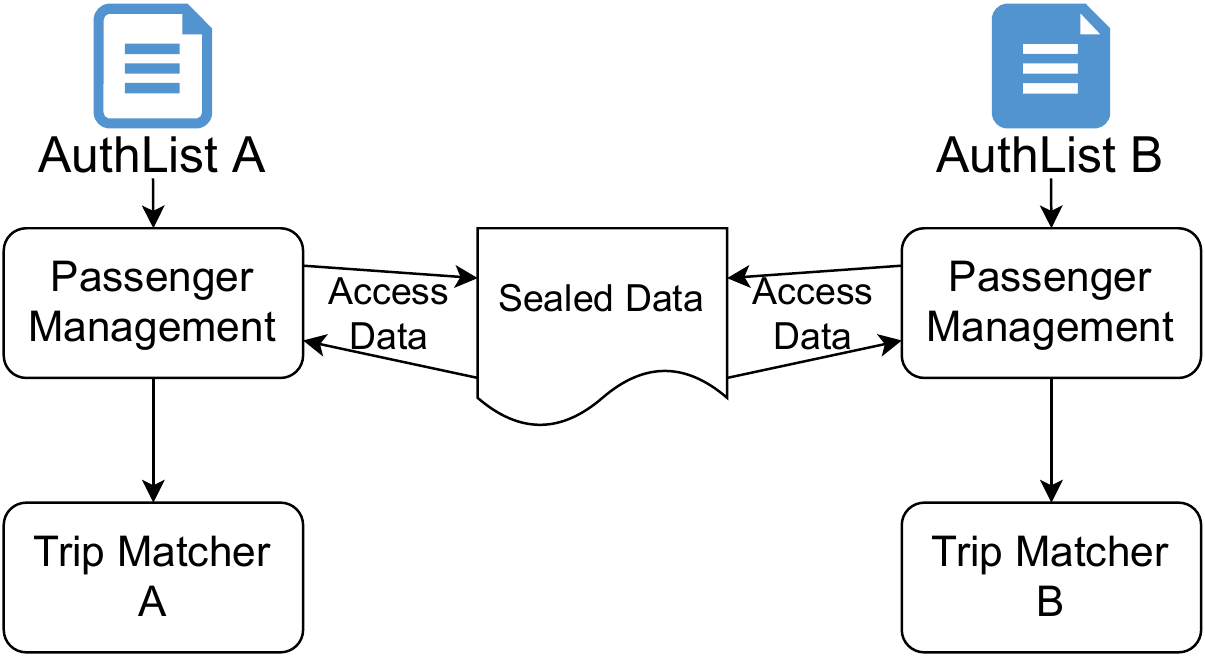}
	\caption{Data sealing with \texttt{MRENCLAVE} scheme only}
	\label{fig:DataSeal}
\end{figure}

Figure~\ref{fig:DataSeal} shows an example of the attack utilizing the
traditional data sealing scheme.
Two instances of passenger management service are generating sealed data and
writing to the permanent drive.
Since their enclave hashes are the same, they can un-seal each other's sealed
data.
This is generally fine for traditional single component enclave application,
because both instances should have the same behavior - if the instance on the
left does not violate confidentiality of user's data, the one on the right
will not violate either.
However, this is not true in distributed applications with multiple enclave
components.
For example, in Figure~\ref{fig:DataSeal}, the instance on the left loads an
\AuthList containing \emph{TripMatcher A}, while the instance on the right
loads \AuthList containing malicious\emph{TripMatcher B}.
Thus, data sealed by the instance on the left could be un-sealed and processed
by the instance on the right and leaked to the malicious TripMatcher.
Therefore, \texttt{MRENCLAVE} scheme is insecure for distributed enclave
applications, because the hash of one enclave does not represent the behavior
of the entire application as a whole.

The \DECENT SDK uses a HMAC-based Key-Derivation Function~\cite{rfc5869} (HKDF)
to derive a key from the \texttt{MRENCLAVE}-derived key
to also bind the key to \emph{a specific \AuthList}.
\DECENT's HKDF scheme prevents malicious hosts from using legitimate
\dComp{s} to leak sealed data to malicious \dComp{s},
and malicious \dComp{s} from directly deriving another \dComp's seal key.

However, binding the \AuthList to seal keys also implies that
sealed data must be explicitly migrated from one application instance to another
if the \AuthList changes or a \dComp is upgraded.
One approach to data migration is to implement a migration API by
which a new \dComp can request sealed data from an existing
\dComp before it is replaced.  For example, if a new \dComp is authorized
by the \dVrfy, it can contact a specified \dComp from which to migrate
data and seal the retrieved data under its own key.
In some scenarios, however, it may be unreasonable to use the existing
\dComp to migrate sealed data. For example, if the \AuthList is changed, or a \dComp is about
to be revoked because of a vulnerability, delaying revocation to
migrate data to a replacement \dComp could subject the application
instance to exploitation.  Guarding against sudden revocation of
a \dComp with a large store of sealed data requires sealing the data
under a key that can be provisioned to ``recovery \dComp{s}'' that
can access and migrate data securely in the event the \dComp's
authority is revoked.

\subsection{Stateless Open Services} \label{ssec:GroupAuthList}

\begin{figure}
	\centering
	\includegraphics[width=0.9\linewidth]{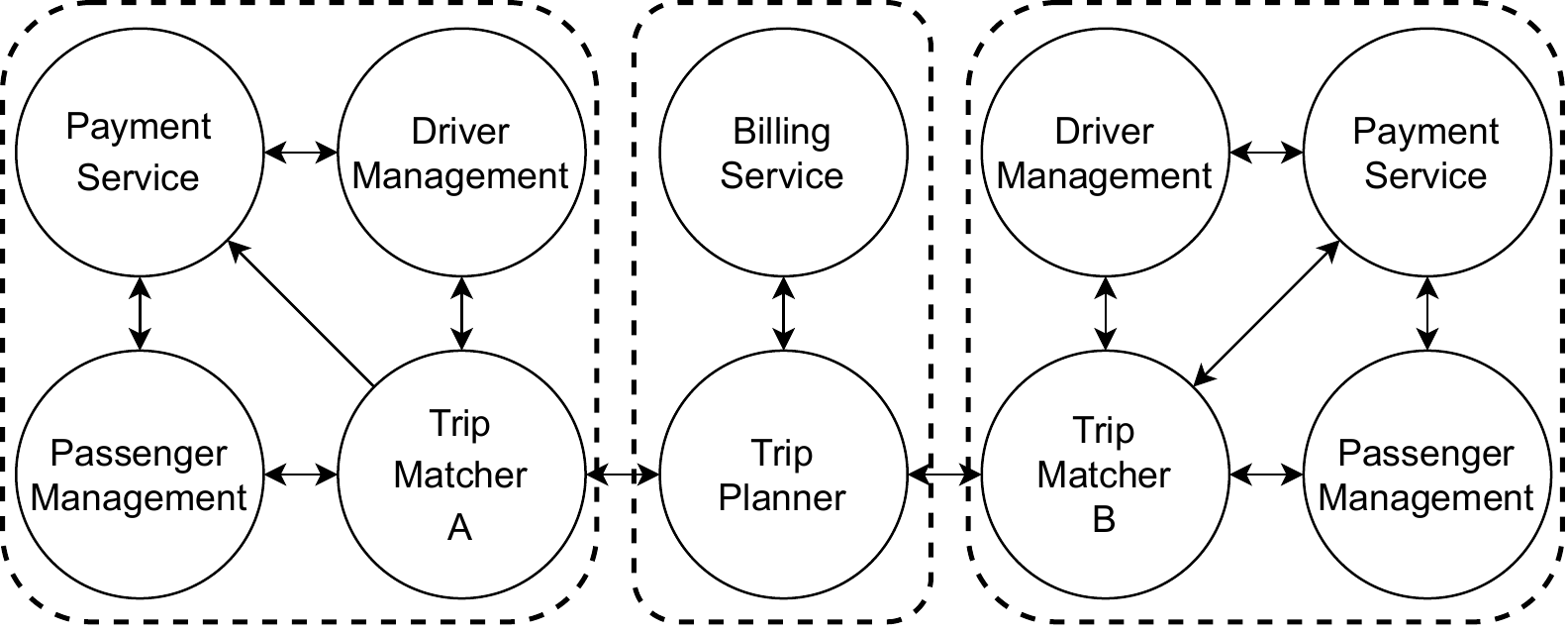}
	\caption{\AuthList with stateless open services}
	\label{fig:AuthListGroup}
\end{figure}

One disadvantage of our approach to component authorization with
\AuthList{s} presented in §\ref{ssec:AuthList} is that component
instances cannot be shared by multiple applications unless their nominal
\AuthList{s} are combined. For example, suppose we wanted to
share the BillingService and TripPlanner components from Figure~\ref{fig:rideshare}
with DecentRide instances that each use a different TripMatcher component,
as illustrated in Figure~\ref{fig:AuthListGroup}.  TripMatcherA does not appear on the
\AuthList of components in the TripMatcherB instance, and vice versa, but both
application instances authorize the same BillingService and TripPlanner components.

Each instance is primarily concerned with protecting the
confidentiality of the user's GPS location and the integrity of the
BillingService's price quote.  An authorized TripPlanner component
with a matching \AuthList ensures the GPS location will not be
revealed to unauthorized entities and that the BillingService could
only be influenced by authorized components.  However, if the
BillingService and TripPlanner are \emph{stateless} with respect to
each incoming request, the TripMatcher only needs to worry about
communication occurring during the processing of its request.  That
is, if no information from TripMatcherA's request is retained after it
has been processed, then none of TripMatcherA's confidential
information can be leaked to TripMatcherB, and TripMatcherB's price
quote cannot be influenced by TripMatcherA's request.
From the perspective of TripPlanner, it only processes the data from the
requester (i.e., TripMatcher) without revealing any additional secret.
Thus, even a malicious trip matcher will not be able to obtain any secret of
TripMatcherA from TripPlanner, or affect the price calculation result returning
to TripMatcherA.
Therefore, TripPlanner does not need to be care about the identity of the
requester.

We propose a simple extension to our \AuthList{s} mechanism that
permits secure sharing of stateless services like the TripPlanner and
BillingService between multiple applications with otherwise
incompatible \AuthList{s}.

\begin{figure}
	\centering

	\begin{subfigure}[t]{0.9\linewidth}
		\centering

		\begin{tabular}{l|l}
			Code Digest           &  Service Definitions   \\ \hline
			\texttt{3fb5...cc46}  &  PaymentService \\
			\texttt{6233...0f6d}  &  TripMatcher   \\
			\texttt{23ed...e470}  &  TripPlanner: \texttt{\{} \\
			                      &  \texttt{~~dff1...8e41: BillingService,} \\
			                      &  \texttt{~~23ed...e470: TripPlanner \}} \\
			\texttt{...}          &  ...
		\end{tabular}

		\caption{Example \AuthList with sub-application}
		\label{sfig:Disc-SubAuthListApp}
	\end{subfigure}
	\vfill
	\begin{subfigure}[t]{0.9\linewidth}
		\centering

		\begin{tabular}{l|l}
			Code Digest           &  Service Definitions   \\ \hline
			\texttt{dff1...8e41}  &  BillingService \\
			\texttt{23ed...e470}  &  TripPlanner \\
		\end{tabular}

		\caption{Example \AuthList for planner \& billing application}
		\label{sfig:Disc-SubAuthListSubApp}
	\end{subfigure}

	\caption{Example nested \AuthList}
	\label{fig:Disc-SubAuthList}
\end{figure}

Figure~\ref{fig:Disc-SubAuthList} demonstrates the \AuthList{s} of
RideShare application (given in Figure~\ref{sfig:Disc-SubAuthListApp})
and planner \& billing sub-application (given in
Figure~\ref{sfig:Disc-SubAuthListSubApp}).  The enclave's hash is not
only mapped to the service name, but also a definition of that
service's components.  The TripPlanner entry given in
Figure~\ref{sfig:Disc-SubAuthListApp} has a nested \AuthList that
defines the authorized components of the TripPlanner sub-application.
These are the components that each TripMatcher application expects to
be in the TripPlanner's \AuthList, which will be confirmed when the
TripMatcher and TripPlanner components establish a secure connection.

\subsection{Dynamic Verification} \label{ssec:AutoVerify}

Currently, dynamic authorization of Decent components requires
stakeholders to separately audit and approve components.  This
introduces some degree of centralization and potential for
exploitation if stakeholders' credentials are compromised or the
stakeholders themselves are malicious.  An alternative we are
currently exploring seeks to shift trust from external auditors onto
automatic verification tools running within Decent.  For example, a
Decent Verifier could formally verify that a new component satisfies a
formal specification of the desired service.  The verification could
be performed on the binary directly or performed on the source and
then compiled (and signed) by the verifier.  While this approach adds
the verification tool chain to the trusted code base, it removes the
ability of external entities to subvert the application with malicious
authorizations, while providing a flexible dynamic authorization
mechanism.

\section{Conclusion} \label{sec:Conclusion}

In this paper, we present the \DECENT Application Platform, a framework
for building secure decentralized applications.
\DecentComp{s} supports mutual authentication without
requiring a universally-trusted entity to authorize \dComp{s}.
\AuthList ensures only authorized \dComp{s} can interact with the system.
\dVrfy{s} and \dRevc{s} allow new \dComp{s} to be authorized or revoked
dynamically.
We formalized \DECENT in ProVerif and verified that
it protects the secrecy and authenticity of application data.

We implemented DecentRide to demonstrate the expressiveness of \DECENT framework,
while the evaluation based on DecentHT shows that, for short sessions,
\DECENT provides 7.5x higher throughput and 3.67x lower latency comparing to
the non-\DECENT implementation.

\ifacknowledgments
\section*{Acknowledgments}
We thank Tuan Tran for feedback on early drafts and
Xiaowei Chu for assistance designing and building DecentHT.  Partial
funding for this research provided by NSF CAREER grant CNS-1750060.
\fi

\newcommand{\showURL}[1]{\unskip}

\ifIsAcmart
	\bibliographystyle{ACM-Reference-Format}
\else
	\bibliographystyle{apacite}
\fi

\bibliography{bibtex/pm-master,bibtex/rfc}

\end{document}